\input harvmac


\def\hat{\widehat}

\def\l{\lambda}

\def\overx{{1\over x-X}}
\def\overy{{1\over y-Y}}

\def\bigvev#1{\left\langle{#1}\right\rangle}
\def\thipi{-{1\over32\pi^2}}
\def\Wsquare{W_\alpha W^\alpha}
\def\Id{1\!\!1}
\def\ZZ{Z\!\!\!Z}

\def\diag{{\rm{diag}}}

\def\a{\alpha}
\def\b{\beta}
\def\t{\widetilde}

\def\o{\overline}

\def\tQ{\tilde Q}


\let\includefigures=\iftrue
\input rotate
\input epsf
\noblackbox
%
%
\includefigures
\message{If you do not have epsf.tex (to include figures),}
\message{change the option at the top of the tex file.}
\def\figin{\epsfcheck\figin}\def\figins{\epsfcheck\figins}
\def\epsfcheck{\ifx\epsfbox\UnDeFiNeD
\message{(NO epsf.tex, FIGURES WILL BE IGNORED)}
\gdef\figin##1{\vskip2in}\gdef\figins##1{\hskip.5in}
\else\message{(FIGURES WILL BE INCLUDED)}%
\gdef\figin##1{##1}\gdef\figins##1{##1}\fi}
\def\DefWarn#1{}

\def\figinsert{\goodbreak\midinsert}
\def\ifig#1#2#3{\DefWarn#1\xdef#1{fig.~\the\figno}
\writedef{#1\leftbracket fig.\noexpand~\the\figno}%
\figinsert\figin{\centerline{#3}}\medskip\centerline{\vbox{\baselineskip12pt
\advance\hsize by -1truein\noindent\footnotefont{\bf
Fig.~\the\figno:} #2}}
\bigskip\endinsert\global\advance\figno by1}
\else
\def\ifig#1#2#3{\xdef#1{fig.~\the\figno}
\writedef{#1\leftbracket fig.\noexpand~\the\figno}%
\global\advance\figno by1} \fi

\def\yboxit#1#2{\vbox{\hrule height #1 \hbox{\vrule width #1
\vbox{#2}\vrule width #1 }\hrule height #1 }}
\def\fillbox#1{\hbox to #1{\vbox to #1{\vfil}\hfil}}
\def\ybox{{\lower 1.3pt \yboxit{0.4pt}{\fillbox{8pt}}\hskip-0.2pt}}



\lref\CDSW{ F.~Cachazo, M.~R.~Douglas, N.~Seiberg and E.~Witten,
JHEP {\bf 0212}, 071 (2002) [arXiv:hep-th/0211170].
}

\lref\CSW{ F.~Cachazo, N.~Seiberg and E.~Witten,
JHEP {\bf 0304},
018 (2003) [arXiv:hep-th/0303207].
}

\lref\okuda{
  T.~Okuda and Y.~Ookouchi,
  arXiv:hep-th/0508189.
}

\lref\zamo{A.~B.~Zamolodchikov,
JETP Lett.\
{\bf 43}, 730 (1986) [Pisma Zh.\ Eksp.\ Teor.\ Fiz.\  {\bf 43},
565 (1986)].
}

\lref\masaki{ C.~h.~Ahn, B.~Feng, Y.~Ookouchi and M.~Shigemori,
Nucl.\ Phys.\ B {\bf 698}, 3 (2004) [arXiv:hep-th/0405101].
}

\lref\DV{R.~Dijkgraaf and C.~Vafa,
Nucl.\ Phys.\ B {\bf
644}, 3 (2002) [arXiv:hep-th/0206255];
Nucl.\ Phys.\ B {\bf 644}, 21 (2002) [arXiv:hep-th/0207106];
arXiv:hep-th/0208048.
}

\lref\ferrari{ F.~Ferrari,
Adv.\ Theor.\ Math.\ Phys.\  {\bf 7}, 619 (2004)
[arXiv:hep-th/0309151].
}

\lref\brodie{ J.~H.~Brodie,
Nucl.\ Phys.\
B {\bf 478}, 123 (1996) [arXiv:hep-th/9605232].
}

\lref\EynardKG{ B.~Eynard,
JHEP {\bf 0301}, 051 (2003) [arXiv:hep-th/0210047].
}

\lref\IW{ K.~Intriligator and B.~Wecht,
Nucl.\ Phys.\ B {\bf 677}, 223 (2004) [arXiv:hep-th/0309201].
}

\lref\CKV{F.~Cachazo, S.~Katz and C.~Vafa,
arXiv:hep-th/0108120.
}

\lref\frashe{ E.~H.~Fradkin and S.~H.~Shenker,
Phys.\ Rev.\ D {\bf
19}, 3682 (1979).
T.~Banks and E.~Rabinovici,
Nucl.\ Phys.\ B {\bf 160}, 349
(1979).
}

\lref\konishi{ K.~Konishi,
Phys.\
Lett.\ B {\bf 135}, 439 (1984).
K.~i.~Konishi and K.~i.~Shizuya,
Nuovo Cim.\ A
{\bf 90}, 111 (1985).
}

\lref\arnold{V.I. Arnold, S.M. Gusein-Zade and A.N. Varchenko,
{\it Singularities of Differentiable Maps} volumes 1 and 2,
Birkhauser (1985), and references therein.}

\lref\farkra{ H. M. Farkas and I. Kra, {\it Riemann Surfaces}, 2nd
ed., 1991, Springer.}

\lref\mio{ L.~Mazzucato,
JHEP {\bf 0411}, 020 (2004) [arXiv:hep-th/0408240].
}

\lref\eynard{ B.~Eynard and J.~Zinn-Justin,
Nucl.\ Phys.\ B {\bf 386}, 558 (1992) [arXiv:hep-th/9204082].
}

\lref\CachazoYC{ F.~Cachazo, N.~Seiberg and E.~Witten,
JHEP {\bf 0304}, 018 (2003) [arXiv:hep-th/0303207].
}

\lref\KSS{
D.~Kutasov,
Phys.\ Lett.\ B {\bf 351}, 230 (1995) [arXiv:hep-th/9503086].
D.~Kutasov and A.~Schwimmer,
Phys.\ Lett.\ B {\bf 354}, 315 (1995) [arXiv:hep-th/9505004].
D.~Kutasov, A.~Schwimmer and N.~Seiberg,
Nucl.\ Phys.\
B {\bf 459}, 455 (1996) [arXiv:hep-th/9510222].
}

\lref\products{
  K.~A.~Intriligator, R.~G.~Leigh and M.~J.~Strassler,
  Nucl.\ Phys.\ B {\bf 456}, 567 (1995)
  [arXiv:hep-th/9506148].
 E.~Barnes, K.~Intriligator, B.~Wecht and J.~Wright,
Nucl.\ Phys.\ B {\bf 716}, 33 (2005) [arXiv:hep-th/0502049].
}

\lref\Cardy{ J.~L.~Cardy,
Phys.\ Lett.\ B {\bf 215}, 749 (1988).
}

\lref\IntriligatorAU{ K.~A.~Intriligator and N.~Seiberg,
Nucl.\ Phys.\ Proc.\ Suppl.\  {\bf 45BC}, 1 (1996)
[arXiv:hep-th/9509066].
}

\lref\klemm{A.~Klemm, K.~Landsteiner, C.~I.~Lazaroiu and
I.~Runkel,
JHEP {\bf 0305}, 066 (2003) [arXiv:hep-th/0303032].
}

\lref\naccu{
C.~I.~Lazaroiu,
JHEP {\bf 0305}, 044 (2003) [arXiv:hep-th/0303008].
S.~G.~Naculich, H.~J.~Schnitzer and N.~Wyllard,
JHEP {\bf 0308}, 021 (2003) [arXiv:hep-th/0303268];
Nucl.\ Phys.\ B {\bf 674}, 37 (2003) [arXiv:hep-th/0305263].
}


\Title{}{\vbox{\centerline{Remarks on the Analytic Structure}
\medskip
\centerline{of Supersymmetric Effective Actions} }}
\smallskip
\centerline{Luca Mazzucato}
\smallskip
\bigskip
\centerline{International School for Advanced Studies
(SISSA/ISAS)} \centerline{\it Via Beirut 2 - 4, 34014 Trieste, and
INFN, sez. di Trieste, Italy}
\medskip

\bigskip
\vskip 1cm

\noindent We study the effective superpotential of ${\cal N}=1$
supersymmetric gauge theories with a mass gap, whose analytic
properties are encoded in an algebraic curve. We propose that the
degree of the curve equals the number of semiclassical branches of
the gauge theory. This is true for supersymmetric QCD with one
adjoint and polynomial superpotential, where the two sheets of its
hyperelliptic curve correspond to the gauge theory pseudoconfining
and higgs branches. We verify this proposal in the new case of
supersymmetric QCD with two adjoints and superpotential
$V(X)+XY^2$, which is the confining phase deformation of the
$D_{n+2}$ SCFT. This theory has three kinds of classical vacua and
its curve is cubic. Each of the three sheets of the curve
corresponds to one of the three semiclassical branches of the
gauge theory. We show that one can continuously interpolate
between these branches by varying the couplings along the moduli
space.

\vskip 0.5cm

\Date{August 2005}


\newsec{Introduction and Summary}

${\cal N}=1$ supersymmetric gauge theories are a natural testing
ground for the study of nonperturbative gauge dynamics. During the
last decade, the tools of holomorphy and symmetries, mainly
developed by Seiberg \IntriligatorAU, made it possible to gain a
deep insight in the strong coupling regime of gauge dynamics. In
theories with a mass gap, there is an alternative method, proposed
by Dijkgraaf and Vafa \DV, that allows to compute systematically
the offshell effective superpotential just above the mass gap,
where the elementary degree of freedom is the glueball. The vacuum
structure of the gauge theory is encoded in an algebraic curve,
which in the original formulation is obtained by the planar limit
of a related matrix model.

Cachazo, Douglas, Seiberg and Witten \CDSW, by studying the ring
of gauge invariant chiral operators, showed that the matrix model
loop equations are reproduced in the gauge theory by a set of
anomalous Ward identities, which are a generalization of the
Konishi anomaly \konishi. They considered in particular an
$U(N_c)$ supersymmetric gauge theory with a chiral superfield $X$
in the adjoint representation of the gauge group, with tree level
superpotential $W=\Tr V(X)$, where $V'(x)$ is a degree $n$
polynomial. The chiral ring of the quantum theory is described by
the hyperelliptic Riemann surface
 \eqn\onecur{y^2=V'(x)^2+\hbar f(x),
 }
where $f(x)$ is a degree $n-1$ polynomial with vanishing classical
limit. This surface is a double--cover of the $x$ plane, which
describes the expectation values of the adjoint $\langle
X\rangle$. The first sheet is visible classically, while the
second one is not accessible semiclassically. In the quantum
theory, the two sheets are connected by $n$ branch cuts and, at
first, the meaning of the ``invisible sheet'' was not clear. Only
when coupling the theory to the chiral superfields in the
fundamental representation it was possible to understand the
nature of the second sheet \CSW. Let us see briefly why,
considering $U(N_c)$ adjoint SQCD with tree level superpotential
 \eqn\treead{
 W=\Tr V(X)+ \t Q m(X) Q,
 }
where $m(x)$ has degree $n-1$ and we suppressed flavor indices. We
have two different classical vacua of this theory. In the
pseudoconfining vacuum the fundamentals vanish and the adjoint has
diagonal expectation values equal to the roots of the adjoint
polynomial $V'(x)$. In the higgs vacuum, also $Q$ and $\t Q$
acquire an expectation value and the adjoint is equal to the roots
of $m(x)$.  The gauge group is generically broken to
$\prod_{i=1}^k U(N_i)$ with $\sum_i N_i=N_c-L$ and $k\leq n$,
where $L$ is the number of higgsed colors, and at low energy the
nonabelian factors confine, leaving a $U(1)^k$ theory. In theories
with fundamentals, once we fix the number of unbroken gauge groups
$k$, there is no order parameter to distinguish the
pseudoconfining and higgs phases in an invariant way \frashe. Thus
one expects that in the full quantum theory the different
classical vacua with the same number of unbroken $U(1)$ gauge
groups can be connected to each other. So we would use the word
{\it branch} rather than phase to label the pseudoconfining and
higgs vacua.

The concept of branches only makes sense in the semiclassical
limit of large expectation values. Following \CSW, we can study
the chiral ring, whose generators are the observables
 \eqn\genecchi{
 M(x)=\t Q{1\over x-X}Q,\qquad T(x)=\Tr{1\over
 x-X}.
 }
These observables are meromorphic functions on the Riemann surface
\onecur, whose only singularities are simple poles. The classical
limit of these operators characterizes the different branches. In
the pseudoconfining branch, $M(x)$ and $T(x)$ are regular on the
first semiclassical sheet, while on the higgs branch these
generators have poles on the first sheet at the higgs eigenvalues
of the adjoint $\langle X\rangle$. We can continuously interpolate
between the two branches by moving the poles between the two
sheets through the branch cuts. Therefore, in this case the first
sheet corresponds to the pseudoconfining branch and the second
sheet to the higgs branch and the connection between classical
phases, or branches, and degree of the curve is clear.

However, more general supersymmetric gauge theories have algebraic
curves of higher degree, which give rise to branched coverings of
the plane with a larger number of sheets. It is not clear what the
meaning of the ``invisible sheets'' is in general.

In this paper, we suggest that this correspondence between the
degree of the curve and the number of branches is a generic
feature of ${\cal N}=1$ theories. Consider a supersymmetric gauge
theory with a matter content such that, once we fix the number of
unbroken $U(1)$s, there is no order parameter to distinguish
between the various classical branches in an invariant way. This
is the case of a theory with fundamentals, for instance. Under
these assumption, we propose that

\vskip0.2cm\noindent {\it An ${\cal N}=1$ supersymmetric gauge
theory with a mass gap is described by a degree $k$ algebraic
curve, where $k$ is the number of different semiclassical branches
of the theory. The curve is a $k$--sheeted covering of the plane,
where each sheet corresponds to a different branch.}\vskip0.2cm

Note that we exclude the case in which the theory has a Coulomb
branch, as it happens in the theory \treead\ for $n=N$. In this
case in fact there is no mass gap.

\subsec{SQCD with Two Adjoints}

As a check of our proposal, in this paper we generalize the
analysis of \CSW\ to the case of an $SU(N_c)$ SQCD with two
adjoint chiral superfields $X$ and $Y$ and a confining phase
superpotential
\eqn\treesupala{W=\Tr V(X)+\l \Tr XY^2+Q \,m(X)  Q.}
We will find that this theory has three kinds of vacua, the
pseudoconfining, the usual ``abelian'' higgs and a new
branch that we will denote ``nonabelian higgs phase''. The
pseudoconfining vacua are the irreps of the equations of motion
with vanishing fundamentals. In the one adjoint case we discussed
above, we have just one dimensional vacua. In this case, a part
from the usual one dimensional vacua $X=a_i\Id$ and $Y=b_i\Id$,
that we will call {\it abelian} vacua, we have also two
dimensional irreps, that we will call {\it nonabelian} vacua, in
which the adjoints are proportional to the Pauli matrices $X=\hat
a_i\sigma_3$ and $Y=d_i\sigma_3+c_i\sigma_1$.\foot{This phenomenon
was first noted in \CKV\ and then discussed in \ferrari\ in the
case of a supersymmetric gauge theory with adjoint fields and no
fundamentals.} The higgs vacua are the ones in which also the
fundamentals acquire an expectation value. First of all, there are
the usual one dimensional higgs vacua, where $X$ and $Y$ are
proportional to the identity, as in the usual abelian vacua. For
this reason, we will denote this vacuum the {\it abelian higgs}
branch. But there is also a new kind of higgs vacuum, the {\it
nonabelian higgs} branch, in which the adjoints are two
dimensional $X=x_h\sigma_3$ and $Y=y_h\sigma_3+y_1\sigma_1$ and
the fundamentals are nonvanishing.

Due to the presence of fundamentals, we expect no phase transition
and in the full quantum theory the three branches will be
connected by continuously varying the couplings. We will study
then the chiral ring in the quantum theory, by means of the DV
method. In order to compute the curve of the gauge theory, we will
use the matrix model loop equations discussed by Ferrari \ferrari,
that in the gauge theory are reproduced by a set of generalized
Konishi anomaly equations. Our analysis confirms that the DV
method works for theories with two adjoint chiral superfields as
well as for one adjoint theories. The gauge theory curve, however,
is not hyperelliptic as in the usual case \onecur, but {\it cubic}
\eqn\cubisu{ y^3+a(x^2)y^2+b(x^2)y+c(x^2)=0,}
where the coefficients are even polynomials depending on the
couplings and the quantum deformations. This curve is the same as
the curve of Ferrari's two matrix model in the planar limit
\ferrari.

To have a clear picture of the phase structure of the quantum
theory, we will consider again the chiral operators $M(x)$ and
$T(x)$ defined in \genecchi. One can solve for these operators by
the method of anomaly equations and find that they are meromorphic
functions on the cubic curve \cubisu, whose only singularities are
simple poles. In particular, the poles of $T(x)$ have integer
residue as in the one adjoint case \CSW. We will show that, by
moving poles between the three sheets, it is possible to connect
continuously all the three branches. Moreover, a natural
correspondence arises between the branches and the sheets: we can
characterize each of the three branches by specifying the sheet on
which $M(z)$ is regular, or by some combination of poles and
residues of $T(x)$. In this way, we verify that our proposed
correspondence between degree of the curve and the number of
branches is satisfied in a very nontrivial way.

\subsec{Outline of the Paper}

The paper is organized as follows. In Section 2 we describe the
classical chiral ring of SQCD with two adjoints and superpotential
\treesupala. There are pseudoconfining vacua of two kinds, one as
well as two dimensional. Correspondingly, we also have two kinds
of higgs vacua, the usual one dimensional and the new two
dimensional ``nonabelian'' higgs vacuum.

In Section 3 we approach the quantum theory by means of the DV
method and study the cubic equation satisfied by the resolvent
$R(x)$. We describe in details the analytic structure of this
gauge theory curve as a three sheeted covering of the plane and
identify the glueballs of the low energy SQCDs as some particular
$A$ periods of the resolvent. In Section 4 we solve for the meson
operators and get a first idea of the correspondence between
branches and number of sheets.

In Section 5 we solve for $T(x)$ and argue that its only
singularities are simple poles with integer residue. We
characterize the three branches by the location of the poles and
show that we can continuously interpolate between the different
classical vacua of the theory with the same unbroken $U(1)$
factors, generalizing the result of \CSW\ to SQCD with two
adjoints. In Section 6 then we discuss our proposal that the
degree of the gauge theory curve is equal to the number of
branches of the semiclassical theory and check it to hold in the
case of SQCD with different extra matter. We also comment on the
truncation of the chiral ring for $n$ even.

The SQCD with two adjoints has an equivalent Seiberg dual
description proposed by Brodie in the superconformal case \brodie.
In Section 7 we generalize this duality to the theory with the
confining phase superpotential \treesupala, in which case we have
one as well as two dimensional pseudoconfining vacua. We will find
the classical duality map, in the spirit of KSS, and, in a simple
case, also the map in the quantum theory, following \mio.

Finally, in Section 8 we suggest some other examples in which to
test our proposed correspondence between branches and sheets. We
present also some speculations about two adjoint SQCD with $E_n$
type superpotentials and its geometric realization.

There are a bunch of Appendices. In the first Appendix we discuss
the generalized Konishi anomalies we used in the main text to
solve for the chiral ring. In Appendix B we review the solution of
cubic algebraic equations. Finally, in Appendix C we find a basis
for the holomorphic differentials on the cubic curve \cubisu\ of
the gauge theory.

\newsec{The Classical Theory}

In this section we study the classical vacua of the theory. We
have the usual pseudoconfining vacua, with vanishing fundamentals,
and we distinguish them in abelian ones, that is one dimensional
irreps of the algebra of the equations of motion, and nonabelian
ones, denoting two dimensional irreps. Then we have the abelian
higgs vacua and a new classical phase that we will call nonabelian
higgs vacuum. The theory is different depending on whether $n$ is
odd or even. In the following we will consider in details the
former case. In the latter, the pseudoconfining
vacua are still one and two dimensional only, but the chiral ring
is not truncated. The analysis of the quantum theory goes through
for both cases with analogous treatments.

Consider an ${\cal N}=1$ supersymmetric $SU(N_c)$ gauge theory
with matter content consisting in two chiral superfields $X$ and
$Y$ in the adjoint representation, $N_f$ fundamentals $Q^f$ and
$N_f$ anti-fundamentals $\t Q_{\t f}$ ($f$ and $\t f$ are the
flavor indices). We let this theory flow to its IR fixed point and
then we turn on the following superpotential
\eqn\treesupal{W=\Tr V(X)+\l \Tr XY^2+\alpha\Tr Y +\t Q \,m(X)
Q.}
where we suppressed flavor indices and we introduced the adjoint
polynomial
 \eqn\coeffmg{\eqalign{
 V(x)&=\sum_{k=1}^n {t_k\over k+1}\,  x^{k+1}+\beta\, x, }}
and  the meson deformation $m(x)=m_1+m_2\,x$ is diagonal in the
flavor indices, while $\alpha$ and $\beta$ are two Lagrange
multipliers enforcing the tracelessness condition.\foot{The
superpotential \coeffmg\ would be irrelevant in the UV for $n>2$,
however there always exists a range of flavors $N_f$ such that it
is a relevant deformation of the IR fixed point \IW\okuda.} For
ease of notations we included the multiplier $\beta$ as a linear
term in the adjoint polynomial \coeffmg. It will be convenient in
the following to separate the odd and even part of the derivative
of this polynomial as $V'(x)=-v_+(x^2)-xv_-(x^2)$. As a result, we
have cast the multiplier $\beta$ into the definition of
$v_+(x^2)$. The equations of motion are
\eqn\eomshi{\eqalign{ &V'(X) +\lambda Y^2+m_2 \t Q Q=0,\cr
&\lambda\{X,Y\}+\alpha=0,}} \eqn\fundah{\t Q m(X)=0,\quad
m(X)Q=0.}
The \eomshi\ are the $X$ and $Y$ equations of motion, while
\fundah\ are the equations for the fundamentals. In addition, by
varying \treesupal\ with respect to the Lagrange multipliers we
get the tracelessness condition $\Tr\,X=\Tr\,Y=0$.

\subsec{Pseudoconfining Vacua}

We consider at first the pseudoconfining vacua, in which the
fundamentals vanish. We want to study the irreducible
representations of the algebra defined by the adjoint equations of
motion \eomshi\ for $\langle Q\rangle=\langle\tQ\rangle=0$. The
Casimirs are $X^2=x^2\Id$, $Y^2=y^2\Id$. Then the first equation
reads $\l y^2=v_+(x^2)+Xv_-(x^2)$ and we can outline two different
cases.

\item{i)} {\it abelian vacua}

The one--dimensional representations are the solutions to
\eqn\onevacual{\cases{y=-{\alpha\over2\l x},\cr \l y^2+V'(x)=0.}}
Thus we have $n+2$ vacua
\eqn\Xveval{
 \langle X\rangle = \pmatrix{
a_1&&&&&\cr &  .&&&&\cr &&a_2&&\cr &&&. &\cr &&&&.},\qquad \langle
Y\rangle = \pmatrix{ b_1&&&&&\cr &  .&&&&\cr &&b_2&&\cr &&&. &\cr
&&&&.} }
where the $X$ expectation values $a_i$ are the roots of
the degree $n+2$ abelian polynomial
\eqn\onepola{ p(x)=x^2\,V'(x)+{\a^2\over 4\l}=0, }
and $b_i=-{\alpha\over 2\lambda a_i}$. Each $a_i,b_i$ has
multiplicity $N_i$ such that $\sum_{i=1}^{n+2}N_i=N_c$. By
imposing the tracelessness condition on the abelian vacua \Xveval\
we fix the multipliers $\alpha$ and $\beta$. The symmetry breaking
pattern is $SU(N_c)\rightarrow
U(1)^{n+1}\times\prod_{i=1}^{n+2}SU(N_{i})$.

\item{ii)} {\it nonabelian vacua}

The only higher dimensional irreps are two dimensional ones, that
we parameterize in terms of the Pauli matrices $X=\hat
a_i\sigma_3$ and $Y=c_i\sigma_1+d_i\sigma_3$. To satisfy the $X$
equation of motion, the odd part of the adjoint polynomial must
vanish, so we have ${n-1\over2}$ nonabelian vacua $\hat a_i$ which
satisfy $v_-(\hat a_i^2)=0$. The $Y$ expectation values are
$d_i=-{\alpha\over2\l}{1\over \hat a_i}$ and $c_i=[\l^{-1}v_+(\hat
a_i^2)-d_i^2]^{1\over2}$. The nonabelian vacua display a $\ZZ_2$
symmetry that acts by reflection of the eigenvalues around the
origin. Note also that $x=0$ is not a solution. Consider the gauge
symmetry breaking in the nonabelian vacua, for simplicity consider
unbroken gauge group $SU(N_c)$ with $N_c$ even. The generic
nonabelian vacuum is given by
\eqn\Xvevtwo{
 \langle X\rangle = \pmatrix{
 \hat a_1 \sigma_3&&&&\cr
&                 .&&&\cr &&\hat a_{2}\sigma_3&&\cr &&&. &\cr
&&&&.\cr},\qquad
 \langle Y\rangle = \pmatrix{
 c_1 \sigma_1+d_1\sigma_3&&&&\cr
&                 .&&&\cr &&c_2 \sigma_1+d_2\sigma_3&&\cr &&&.
&\cr &&&&.\cr}, }
where each $\hat a_i$ has multiplicity $\hat N_i$ such that
$2\sum_{i=1}^{n-1\over2}\hat N_i=N_c$. In this case, unlike the
usual one--dimensional one, the vacuum decreases the rank of the
gauge group. The gauge symmetry is broken as $SU(N_c)\rightarrow
U(1)^{n-1\over2}\times\prod_{i=1}^{n-1\over 2}SU(\hat N_i)$.

One can easily show that there are no higher dimensional irreps of
the equations of motion \eomshi, following \CKV. One can shift
$X\to X+aY$ and $Y\to Y+bX$ and get to a new algebra with
$X^2=Y^2=0$ and $\{X,Y\}+c=0$. This algebra has just one
irreducible representation, which is two dimensional and
corresponds to the Fock space of a single fermionic
creation--annihilation algebra.\foot{This argument holds
irrespectively of $n$, so we have a finite number of vacua both if
$n$ is odd and even.} The generic gauge symmetry breaking pattern,
in the pseudoconfining case, is the following
\eqn\genbre{ SU(N_c)\longrightarrow
U(1)^{{3\over2}(n+1)-1}\times\prod_{i=1}^{n+2}SU(N_{i})\times
\prod_{i=1}^{n-1\over 2}SU(\hat N_i), }
where $N_c=\sum_{i=1}^{n+2}N_i+2\sum_{i=1}^{n-1\over2}\hat N_i$.
At energies below the vevs but above the dynamical scale of the
theory, we flow to a bunch of ${3\over2}(n+1)$ low energy SQCDs
with massive fundamentals, whose number is $N_f$ in the $n+2$
abelian vacua and $2N_f$ in the $(n-1)/2$ nonabelian vacua. At low
energies we are left with a $U(1)^{{3\over2}(n+1)-1}$ theory.

\subsec{Higgs Vacua}

The equations of motion \eomshi\ allow also for higgs solutions,
in which the fundamentals acquire a vacuum expectation value. The
Yukawa coupling contains just terms in the dressed $X$--mesons.
There are two different kinds of higgs solutions. The first one is
the usual one dimensional vacuum, that we will denote abelian
higgs, but it turns out that there are also new two dimensional
solutions, analogous to \Xvevtwo, that we will denote nonabelian
higgs. We consider for simplicity the higgsing of just the last
flavor.

\item{i)} {\it abelian higgs}

The usual one--dimensional higgs vacua are given by
\eqn\quahi{\eqalign{ \t Q_{N_f}=(\t h,0,\ldots,0),&\quad
Q^{N_f}=(h,0\ldots,0),\cr X=\diag(x_h,{\, pseudoconf.}),&\quad
Y=\diag(y_h,{\, pseudoconf.}), }}
where $x_h$ is a root of the meson deformation $m(x)$, i.e.
$x_h=-m_1/m_2$, and $y_h=-{\alpha\over2\lambda x_h}$ and the
squark expectation values are fixed by the $X$ equations of motion
to $\t h h=-{1\over m_2}[V'(x_h)+\lambda y_h^2]$. The remaining
diagonal expectation values for the adjoints in \quahi\ are the
generic pseudoconfining vacua \Xveval\ and \Xvevtwo\ and we have
to impose the tracelessness. The symmetry breaking pattern is as
in \genbre\ but now the sum of the ranks of the low energy SQCDs
decreases by one, $\sum_{i=1}^{n+2}N_i+2\sum_{i=1}^{n-1\over2}\hat
N_i=N_c-1$. In this abelian higgs case we higgs one color
direction.

\item{ii)} {\it nonabelian higgs}

The equations of motion \eomshi\ admit also two--dimensional
representations with nonvanishing fundamentals
\eqn\nonhi{ \eqalign{ \t Q_{N_f}=(\widehat{\t
h},0,\ldots,0),&\quad Q^{N_f}=(\widehat h,0\ldots,0),\cr
X=\diag(x_h \sigma_3,{\, pseudoconf.}),&\quad Y=\diag(\hat
y_1\sigma_1+ y_h\sigma_3,{\, pseudoconf.}), } }
where $x_h$ is always a root of the meson deformation $m(x)$,
while
$$
y_h=-{\alpha\over2\lambda x_h},\qquad \lambda \hat
y_1^2+V'(-x_h)+{\alpha^2\over4\lambda x_h^2}=0,
$$
and the quark expectation values are $\widehat{\t h}\hat
h=-{1\over m_2}[V'(x_h)-V'(-x_h)]$. Again, the remaining diagonal
expectation values for the adjoints in \nonhi\ are the generic
pseudoconfining vacua \Xveval\ and \Xvevtwo\ and we have to impose
the tracelessness. The symmetry breaking pattern is as in \genbre\
but now the sum of the ranks of the low energy SQCDs decreases by
two, $\sum_{i=1}^{n+2}N_i+2\sum_{i=1}^{n-1\over2}\hat N_i=N_c-2$.
In this nonabelian higgs case we higgs {\it two} color directions
and this represents a new classical phase of SQCD.

\subsec{D--terms}

Consider the kinetic term for the adjoints
$$\int d^2\theta\,d^2\bar\theta\,\Tr\left(X^\dagger e^{ad V}X+ Y^\dagger
e^{ad V}Y\right),
$$
the D--term equations of motion are
$[X,X^{\dagger}]+[Y,Y^\dagger]=0$. The abelian vacua \Xveval,
satisfy the D--term equation as usual. For the nonabelian vacua
\Xvevtwo\ and \nonhi, however, due to the nonvanishing commutator
of the Pauli matrices, we get the additional condition
\eqn\dcond{ \matrix{{\rm pseudoconf.}\qquad&\qquad{\rm
nonabelian\,\,higgs}\cr&&\cr {\rm Im}\,\, c
d^*=0,\qquad&\qquad{\rm Im}\,\, \hat y_1 y_h^*=0.} }
Note also that, if we set to zero the Lagrange multiplier
$\alpha$, then the term proportional to $\sigma_3$ in $\langle
Y\rangle$ vanishes, so that the nonabelian vacuum automatically
satisfies the D--term. This would amount to consider $Y$
transforming in the adjoint of $U(N_c)$, rather than $SU(N_c)$. In
this way we would get rid of this additional D--term condition,
since the vev that is subject to the constraint is proportional to
$\alpha$. However, if we compute the low energy matter content in
the nonabelian vacua \Xvevtwo, we find that the $\Tr Y$, which is
the $U(1)$ part of the adjoint, becomes massless in this case.
Albeit being neutral under the gauge interactions, the $\Tr Y$
field interacts with the other massive low energy degrees of
freedom through superpotential terms. On the other hand, we need a
mass gap in order to make sense of the glueball superpotential, so
we are forced to keep the Lagrange multiplier $\alpha$ and the
additional constraint \dcond\ and we will consider the $SU(N_c)$
gauge theory in the following.

\subsec{The Classical Chiral Ring}

Consider the superpotential \treesupal\ and for simplicity drop
all the lower relevant operators, keeping just $V'(x)=t_n x^n$. In
this case the theory is superconformal and its flows have been
studied in \IW. Using the equations of motion we get
$\left((-)^n+1\right)X^nY=-2 Y^3$, so that in the $n$ odd case the
chiral ring is truncated to $Y^3=Y$ and is generated by the
products $\Tr X^{k-1}Y^{j-1}$, for $k=1,\ldots,n$ and $j=1,2,3$,
regardless of the ordering. Due to $\{X,Y\}=-{\alpha\over
\lambda}$ and the cyclicity of the trace, the only nonvanishing
chiral ring operators are actually
\eqn\chicl{\eqalign{ &\Tr X^{k-1},\quad k=3,\ldots,n,\qquad\Tr
Y^2,\cr &\Tr X^{2k} Y^2,\quad k=1,\ldots,{1\over2}(n-1),}}
and also the dressed mesons $M_{kj}=\t Q X^{k-1}Y^{j-1}Q$, for
$k=1,\ldots,n$ and $j=1,2,3$.

In the $n$ even case, apparently, the chiral ring would not be
truncated. We will consider just the $n$ odd case in the following
and get back to this issue in Section 5.3, where we will show that
indeed, by considering the flow from $n$ odd to $n'$ even, with
$n'<n$, the chiral ring is truncated also in the even case.

We will be interested in solving for the expectation values of the
operators of the chiral ring. We can collect them in four
generating functions
\eqn\chiralgen{\eqalign{ Z(x,y)&=-{1\over32\pi^2}\bigvev{\Tr
{W_{\alpha}W^{\alpha}\over x-X}{1\over y-Y}},\cr
u_{\alpha}(x,y)&={1\over 4\pi}\bigvev{\Tr {W_{\alpha}\over
x-X}{1\over y-Y}},\cr U(x,y)&=\bigvev{\Tr{1\over x-X}{1\over
y-Y}},\cr M_{\t f}^f(x,y) &= \bigvev{\t Q_{\t f} {1\over
x-X}{1\over y-Y} Q^f}. }}
In a supersymmetric vacuum $u_\alpha$ must be vanishing, therefore
we set it to zero. These loop functions \chiralgen\ can be
expanded in Laurent series of $x$ or $y$, for instance the first
one is
\eqn\Lau{Z(x,y)= \sum_{k=0}^\infty
x^{-1-k}R_k^Y(y)=\sum_{k=0}^\infty y^{-1-k}R_k^X(x),}
where we introduced the {\it generalized resolvents}
\eqn\genres{R_k^X(x)=\thipi\bigvev{\Tr{\Wsquare\over x-X}Y^k} ,}
and analogously for $R_k^Y(y)$. The leading term in the expansion
\Lau\ is the usual resolvent of the one--adjoint theory. It will
be useful to introduce also a {\it generalized glueball} $\t
S=\thipi\langle\Tr\Wsquare Y\rangle$. Since all the single trace
operators of two adjoints can be extracted from $R_k^X(x)$, we can
just solve for this operators and do not consider $R^Y_k(y)$.
Analogous expressions to \genres\ hold for the other generalized
resolvents
\eqn\gemete{ M_k(x)=\bigvev{\t Q{1\over x-X}Y^kQ},\qquad
T_k(x)=\bigvev{\Tr {1\over x-X}Y^k}. }

Let us consider the semiclassical expressions for the generators
that we obtain by plugging into \chiralgen\ the solutions for
$\langle X\rangle$ and $\langle Y\rangle$ and for the
fundamentals. Classically, the glueball vanishes, however we can
keep it as a fixed parameter to study $Z(x,y)$. Then, we can
Laurent expand it and obtain the semiclassical resolvent
\eqn\Rabal{ R(x)=\sum_{i=1}^{n+2}{S_{i}\over x-a_i}+
2x\sum_{i=1}^{n-1\over2}{\hat S_i\over x^2-\hat a_i^2}, }
which has poles at the classical vacua. The residues are $S_i$,
for $i=1,\ldots,n+2$, the glueballs for the SQCDs we flow to in
the abelian vacua \Xveval, and $\hat S_i$, for
$i=1,\ldots,(n-1)/2$ the glueballs for the SQCDs we flow to in the
nonabelian vacua \Xvevtwo.

The meson generator depends on the vacuum we consider. In the
pseudoconfining phase $M(x,y)$ vanish, since the fundamentals
vanish. In the abelian higgs \quahi\ and nonabelian higgs phases
\nonhi, however, it is nonvanishing and we find
\eqn\meshig{\eqalign{ {\rm abelian
higgs}\qquad\qquad\qquad&\qquad\qquad\qquad{\rm nonabelian
higgs}\cr M(x)=-{1\over m'(x_h)}{V'(x_h)+{\alpha^2\over4\lambda
x_h^2}\over x-x_h},\qquad &M(x)=-{1\over m'(x_h)}{V'(x_h) -
V'(-x_h)\over x-x_h}, }}
The meson generator has poles at the higgs eigenvalues, whose
residue depends on the couplings and the branch.\foot{We
considered just the last flavor direction, i.e.
$M(x)=M(x)^{N_f}_{N_f}$, according to the classical solutions
\quahi\ and \nonhi.}

\newsec{The Three--sheeted Curve}

In this Section we will study the chiral ring in the quantum
theory by making use of the Konishi anomaly equations. We will
consider in details the curve of the gauge theory \treesupal,
defined by the gauge theory resolvent $R(x)$. On the other hand,
due to the DV correspondence, this curve provides the solution to
the planar limit of the two--matrix model, whose action is given
by the gauge theory superpotential \treesupal. This matrix model
has been solved by Ferrari \ferrari\ and we will collect its basic
features in Appendix A, in a gauge theory language.

In the following, we will study the semiclassical expansion of the
resolvent and identify its analytic structure, i.e. the branch
points. In Appendix C we will work out the holomorphic
differentials on the curve. Let us briefly recall that in the
one--adjoint theory, described in \CSW, the curve is the
hyperelliptic Riemann surface $w^2=V'(x)^2+\hbar f(x)$. Each of
the two sheets corresponds to a classical phase of the gauge
theory, that are the pseudoconfining and the higgs branches. The
interpolation between the two branches is possible by continuously
move the poles of the resolvents $M(x)$ and $T(x)$ through the two
sheets. In our two adjoint theory, we have instead a cubic
algebraic curve and the full quantum theory is described by a
three--sheeted covering of the plane. We will see in the next
Sections that again each of the three sheets corresponds to one
different branch, i.e. the pseudoconfining, the abelian higgs and
the nonabelian higgs branches. This will confirm the proposal that
the degree of the ${\cal N}=1$ curve corresponds to the number of
semiclassical branches of the gauge theory. In the quantum theory
we can interpolate between all the phases by moving poles around
the curve.

\subsec{The Cubic Equation}

Unlike the one--adjoint theory \CDSW, in this more general case
there is some work to do in order to extract the algebraic
equation of the curve from the anomaly equations. The strategy we
will use is to derive some equations involving the generator
$Z(x,y)$ in \chiralgen\ and then, by considering the Laurent
expansion of these equations, derive some recursion relations for
$R_k(x)$ that magically close on the resolvent $R(x)$, which
defines the algebraic curve. This procedure, proposed by Ferrari
\ferrari\ in the corresponding matrix model, is reproduced as well
on the gauge theory side. We summarize the basic Ward identities
in Appendix A.

The curve of the gauge theory is the following cubic
\eqn\cubbiw{ w^3+a(x^2)w^2+b(x^2)w+c(x^2)=0, }
where we introduced the auxiliary variable
$w=x^2\left[R(x)-V'(x)\right]$. The coefficients are given by
\eqn\abici{\cases{ \eqalign{
a(x^2)=x^2[V'(x)+V'(-x)]-{\a^2\over4\l} },\cr\cr \eqalign{
b(x^2)=x^4V'(x)V'(-x)-{\a^2 x^2\over4\l}[V'(x)+V'(-x)]+\hbar
x^2\left[x^2[F_0(x)+F_0(-x)]+\a \t S\right], }\cr \eqalign{
c(x^2)=&x^4\Bigl\{-{\a^2\over4\l}V'(x)V'(-x)+\hbar\Bigl[x^2\bigl[F_0(-x)V'(x)+F_0(x)V'(-x)+2\l\t
F_2(x^2)+\a\t F_1(x^2)\bigr]\cr &+{\a\over2}\t
S[V'(x)+V'(-x)]\Bigr]-\hbar^2\lambda\t S^2 \Bigr\}. } }}
Let us clarify the notations. We have introduced the degree $n-1$
polynomials $F_k(x)$
\eqn\delte{ F_k(x)\equiv\thipi\bigvev{\Tr\Wsquare
{V'(x)-V'(X)\over x-X}Y^k}. }
These are the generalization of the usual quantum deformation
$f(x)$ in the one--adjoint hyperelliptic Riemann surface
$w^2=V'(x)^2+f(x)$ and their coefficients are proportional to the
glueballs, hence they vanish classically.\foot{Our notations are
slightly different from those of CDSW \CDSW, i.e.
$f(x)=-4F_0(x)$.} Moreover, we introduced the even combinations
$2x^2\t F_1(x^2)=x[F_1(x)-F_1(-x)]$ and $2\t
F_2(x^2)=F_2(x)+F_2(-x)$. Note that the coefficient of the leading
term of the first polynomial $F_0(x)$ is the glueball
$S=F_{0(n-1)}/t_n$.

The polynomial coefficients $a,b,c$ of the curve \cubbiw\ are even
functions of $x$. The curve in fact is invariant under the
automorphism $x\to-x$, that we will discuss below.

As explained in Appendix B, it is convenient to shift $w\to
w+a(x)/3$ to get rid of the subleading term and cast \cubbiw\ to
its normal form
\eqn\bibi{ f(w,x)=w^3+3\gamma(x^2)w+2\delta(x^2)=0, }
where $\gamma(x^2)$ and $\delta(x^2)$ are the combinations
$3\gamma=(b-{a^2\over3})$ and
$2\delta=(c-{ab\over3}+{2a^3\over27})$ of \abici. Let us introduce
the discriminant of the cubic equation
$\Delta(x^2)=\gamma^3+\delta^2$ and the auxiliary function
$u(x)=(-\delta+\sqrt{\Delta})^{1\over3}$. The general solutions to
\cubbiw\ is given by
\eqn\solut{\eqalign{
w^{(I)}=&e^{i{2\over3}\pi}u-e^{-i{2\over3}\pi}{\gamma\over
u}-{a\over3},\cr w^{(II)}=&u-{\gamma\over u}-{a\over3},\cr
w^{(III)}=&e^{-i{2\over3}\pi}u-e^{i{2\over3}\pi}{\gamma\over
u}-{a\over3}, }}
and recalling the definition $w=x^2[R(x)-V'(x)]$ we find the three
expressions for the resolvents
$$
\hbar R^{(i)}(x)=V'(x)+w^{(i)}(x)/x^2,\qquad i=I,II,III.
$$
The resolvent on the physical sheet will be identified as usual by
its asymptotics $R(x)\sim S/x$. We can rearrange the asymptotic
expansion as a semiclassical expansion in powers of $\hbar$ and
find that the solution $w^{(I)}(x)$ has the correct physical
behavior
\eqn\semifi{R(x)={x^2F_0(x)+{\alpha\t S\over2}\over
x^2V'(x)+{\a^2\over4\l}}+\l x{2\t F_2(x^2)+\alpha\t
F_1(x^2)\over2v_-(x^2)[x^2V'(x)+{\a^2\over4\l}]}+{\cal O}(\hbar),
}
where $-2xv_-(x^2)=V'(x)-V'(-x)$ is the odd part of the adjoint
polynomial and  $S=F_{0(n-1)}/t_n$. The other two solutions in
\solut\ describe the second and the third sheet, which are not
visible classically, and we collect them in Appendix A. Their
leading term in the semiclassical expansion is
\eqn\semialli{\eqalign{ \hbar
R^{II}(x)=&V'(x)+{\alpha^2\over4\lambda x^2}+{\cal O}(\hbar),\cr
\hbar R^{III}(x)=&V'(x)-V'(-x)+{\cal O}(\hbar). }}
If we put $\hbar =0$ in the anomaly equations, but still keeping
the glueballs as parameters, $F_k(x)\neq0$, we find as classical
expression for the resolvent precisely the physical solution
\semifi.

\subsec{The Branch Points}

Let us look at the analytic structure of the curve
\bibi. The branch points are the singular points of the
curve, that is the points at which $f=df=0$. Since $\partial_w
f=3(w+\gamma)$, one can easily find that the singular points are
given by the zeros of the discriminant
$\Delta(x^2)=\gamma^3+\delta^2$. The ramification index $r_i$ of
each of these branch points is such that $f(w,x)$ together with
its $r_i-1$ derivatives vanish at the point. This index tells how
many sheets we can reach by winding around the branching point.
The number of branch points would be $\deg\Delta=6(n+2)$ where $n$
is the degree of $V'(x)$ in the gauge theory. However, we can
collect out an overall $x^6$ factor in front of $\Delta$.
Therefore, the number of branch points is $6(n+1)$. All of these
branching points have ramification index $r_i=2$ since
$\partial^2_w f(w,x)=6w$ never vanishes at these points.

Let us look at the semiclassical expansion of the discriminant
\eqn\discricla{ -(27/x^6)\Delta=v_-(x^2)^2 p(x)^2p(-x)^2 +{\cal
O}(\hbar). }
Classically, we have $6(n+1)$ double zeros, which come in pairs
symmetric under $x\to-x$. The first $n-1$ of them come from the
zeros of $v_-(x^2)$, the odd part of the adjoint polynomial, and
correspond to the nonabelian vacua \Xvevtwo. Other $n+2$ double
zeros are given by the roots of $p(x)$ in \onepola\ and correspond
to the abelian vacua \Xveval. The last $n+2$ double zeros are
given by the roots of $p(-x)$ and are the reflection of the
abelian vacua \onepola\ under $x\to-x$. The latter are not
classical vacua and we will shortly see how they arise. Consider
now the image of $x=0$. Even if $\Delta$ has an overall factor
$x^6$, it turns out that $\partial_w f$ vanishes at $x=0$ on the
first and third sheet, but it is nonvanishing on the second sheet,
so this is actually a cusp and not a branch point. Each double
zero of the discriminant corresponds to a classical pole in the
semiclassical expansion of the resolvent on some particular
sheets.

In the quantum theory, each of these double zeros split up into
two branch points. Since they all have branching number two, each
branch point lies on two sheets of the algebraic curve. Therefore,
to tell which sheets are connected by which branch point, we have
to solve the conditions
\eqn\fise{ w^{(i)}(a)=w^{(j)}(a),\qquad i,j=I,II,III, }
where $w^{(i)}$ are the three solutions \solut. For instance, the
points $x=a$ that lie on the first and second sheet satisfy
$w^{(I)}(a)=w^{(II)}(a)$, which gives the condition
$(-\delta(a))^{2\over3}=e^{i{\pi\over3}}\gamma(a)$. In practice,
however, the expressions of $\gamma(a)$ and $\delta(a)$ are very
complicated and involve the quantum deformations $F_k(x)$, but it
turns out that it is sufficient to study the limit $\hbar=0$, in
which the cubic \cubbiw\ factorizes into three disconnected sheets
\eqn\classil{
\left[w-{\a^2\over4\l}\right][w^2+x^2[V'(x)+V'(-x)]w+x^4V'(x)V'(-x)]=0,
}
whose solutions we can identify as the semiclassical limits of the
resolvent on the three sheets
\eqn\classol{ w^{(I)}_{cl}(x)=-x^2V'(x),\quad
w^{(II)}_{cl}(x)={\a^2\over4\l},\quad
w^{(III)}_{cl}(x)=-x^2V'(-x). }
In the exact solutions \solut\ they satisfy \fise, which means
that they are special points lying on two different sheets. In the
classical limit, each couple of branch points degenerates into a
pole located at the corresponding vacuum. We can solve the
conditions \fise\ on the classical curve and identify which
classical pole connects which sheets: on the curve \classil\ the
vacua are represented by marked points on each disconnected
sheets, such that the above conditions are satisfied. We find then
\eqn\fisecl{ w^{(I)}_{cl}(a)=w^{(II)}_{cl}(a)\quad iff\quad
a^2V'(a)+{\a^2\over4\l}=0, }
\eqn\fithecl{ w^I_{cl}(a)=w^{III}_{cl}(a)\quad iff\quad
v_-(a^2)=0, }
\eqn\fishecl{ w^{(II)}_{cl}(a)=w^{(III)}_{cl}(a)\quad iff\quad
a^2V'(-a)+{\a^2\over4\l}=0. }
The branch points \fisecl\ connecting the 1st with the 2nd sheet
come from the splitting of the abelian vacua $x=a_i$, for
$i=1,\ldots,n+2$. We call $I_i$ the $n+2$ branch cuts coming from
the splitting of the abelian vacua at $a_i$. These branch cuts
connect the 1st and 2nd sheet. The 1st and the 3rd sheet are
connected by the branch points \fithecl\ coming from the splitting
of the nonabelian vacua $x=\pm\hat a_i$, for $i=1,\ldots,(n-1)/2$.
Note that for each nonabelian vacuum corresponding to a value of
$\hat a_i^2$, we have on the 1st sheet a pole in $\hat a_i$ and
another one in $-\hat a_i$. We denote $\widehat I_i$ the branch
cuts around $\hat a_i$ and $\widehat I'_i$ the branch cuts around
$-\hat a_i$. These branch cuts connect the 1st and 3rd sheets.
Finally, the 2nd and 3rd sheet are connected by the branch points
\fishecl\ that split from $x=-a_i$, which are the reflections of
the abelian vacua $a_i$. We will denote $I_i'$ the branch cut
around the pole at $-a_i$. These cuts connect the 2nd and 3rd
sheet. Note that these cuts do not correspond to a gauge theory
vacuum and, indeed, we cannot see them on the physical sheet. In
this way we have accounted for all the $6(n+1)$ branch points. We
can summarize the monodromy structure of the curve as follows
\eqn\monod{ \matrix{ {\rm Branch\,Cuts} & {\rm Sheets} \cr I_i &
I\leftrightarrow II \cr I_i' & II\leftrightarrow III\cr \widehat
I_i,\widehat I'_i   & I\leftrightarrow III
                }}
We could have found the same results by looking at the
semiclassical expression for the resolvent on the three sheets in
\semifi\ and in Appendix A. In fact, the classical poles in the
resolvents correspond precisely to the marked points on the three
disconnected sheets. In the quantum theory, each marked point
splits into two branch points.

The automorphism $x\to-x$ of the curve \cubbiw\ exchanges the 1st
and the 3rd sheets, while leaving the second sheet invariant. We
do not discuss the noncompact $B$ cycles. For our purpose, in
fact, we will always consider the gauge theory in the weak
coupling expansion, so the periods on compact and noncompact cycle
will never mix.

\ifig\oneloopdiag{The three sheeted curve. The cut $I_i$ comes
from the splitting of the abelian vacua at $x=a_i$. The cuts
$\widehat I_i$ and $\widehat I'_i$ come from the splitting of the
nonabelian vacuum, at $x=\pm \hat a_i$. The cut $I'_i$ is not
visible from the physical sheet. It comes from the splitting of
the pole at $x=-a_i$. The A--periods are also shown.}
{\epsfxsize=0,8\hsize\epsfbox{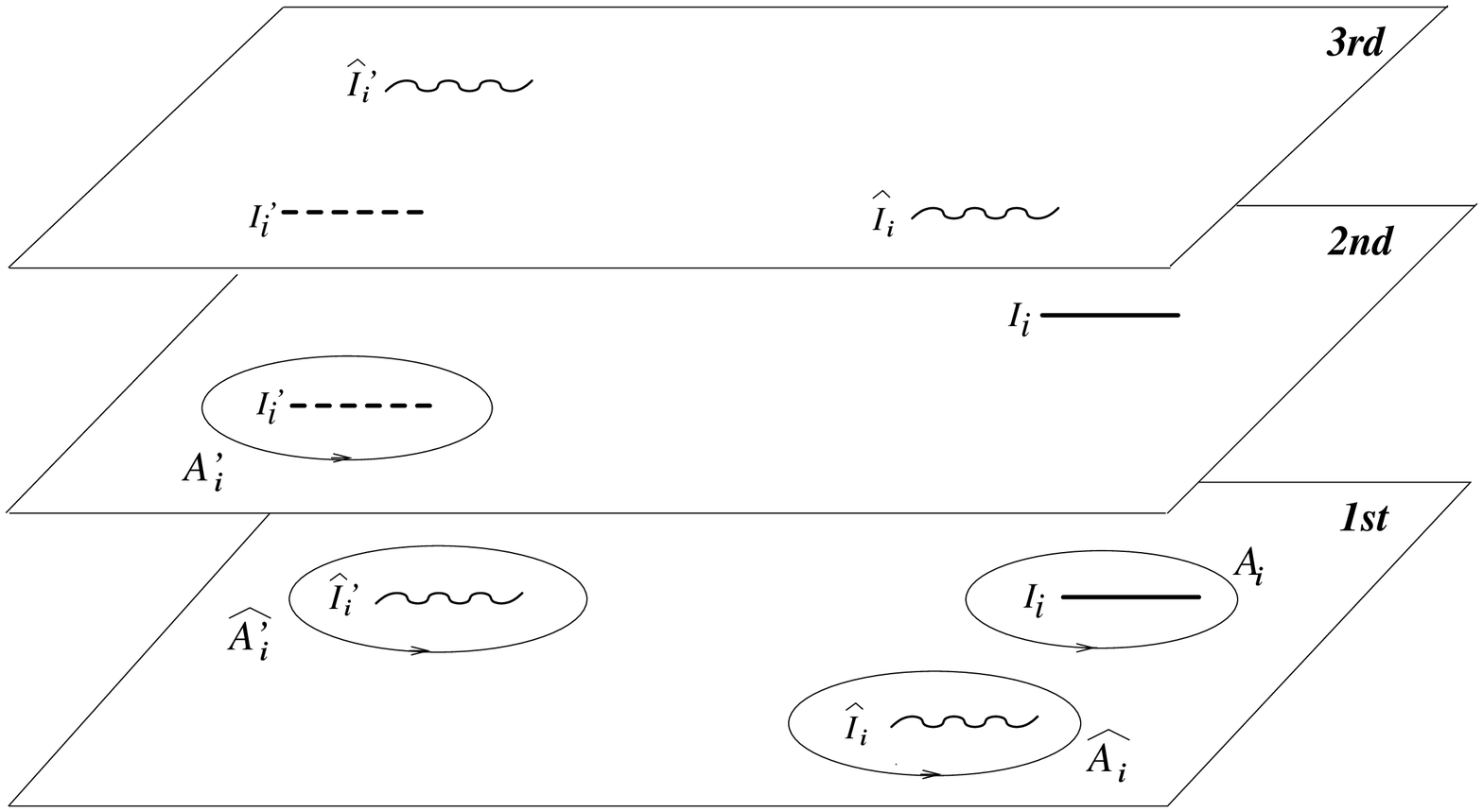}}

Now we can use the Hurwitz formula $2g-2=-2p+\sum_i(r_i-1)$ to
compute the genus $g$ of the curve, where $p$ is the number of
sheets and the sum runs over all the branch points, $r_i=2$ being
the branching number of each. We find
\eqn\genus{ g=3(n+1)-2. }
In Appendix C we compute the holomorphic differentials on the
curve \cubbiw, showing that we have $g$ of them.

\subsec{The Glueballs}

In the one--adjoint theory \CDSW, the usual way one defines the
glueball $S_i$ in the $i$--th low energy SQCD is by computing the
period of the resolvent around the $i$--th cut on the physical
sheet. In our case, the generic low energy SQCDs \genbre\ come
from abelian as well as nonabelian vacua and they require
different definitions. In the case of the abelian vacua \Xveval,
we define the glueballs as usual
\eqn\abeglu{ S_i=\oint_{A_i}R(x), }
where the $A_i$ contour sorrounds the corresponding $I_i$ abelian
cut, see Fig.1. This definition reproduces the semiclassical
result \Rabal\ and is the same prescription as in \CDSW. On the
other hand, the SQCD we flow to in the nonabelian vacuum is
described on the physical sheet by the two cuts $\hat I_i$ and
$\hat I_i'$, which are symmetric with respect to the origin. This
phenomenon has been called "eigenvalue entanglement" in the
related two--matrix model \ferrari, where it was shown that the
eigenvalue density $\rho(x)$ for such representations is
symmetric, $\rho(x)=\rho(-x)$ for $x\in \hat I_i\cup\hat I_i'$.
Since the gauge theory glueball corresponds to the matrix model
filling fraction of the eigenvalues, the periods of the resolvent
$R(x)$ around the cuts $\hat I_i$ and $\hat I_i'$ is the same. We
define therefore the glueball as either period
\eqn\nabeglu{ \hat S_i=\oint_{\hat A_i}R(x)=\oint_{\hat
A_i'}R(x).}
This definition is consistent with the semiclassical resolvent
\Rabal, in fact we have that the total glueball is
$S=\sum_{i=1}^{n+2}S_i+2\sum_{i=1}^{n-1\over2}\hat S_i$ and is the
residue of the resolvent at the pole at infinity. We will see
below that this definition reproduces also the Konishi anomalies
in these low energy SQCDs.

We would like to find that the number of glueballs corresponds to
the number of parameters in the equation for the resolvent
\cubbiw, which in turn is related to the genus of the curve.
Recall first what happens in the one--adjoint theory with gauge
group $U(N)$ \CDSW. There, a degree $n$ adjoint polynomial $V'(x)$
gives $n$ low energy SQCD blocks with gauge group $U(N_i)$, each
of which defines a glueball $S_i$ \CDSW. The $n$ glueballs $S_i$
are in one to one correspondence to the $n$ coefficients of the
quantum deformation $f_{n-1}(x)$ of the ${\cal N}=1$ hyperelliptic
curve $y^2=V'(x)^2+f_{n-1}(x)$. Finally, we can fix the
coefficient of the leading term of $f(x)$ by the residue of the
resolvent at infinity, due to the overall relation $\sum_i S_i=S$.
The number of moduli of the curve is just the genus $g=n-1$, and
the free parameters in $f(x)$ actually parameterize the moduli of
the curve.

Now let us look at the cubic curve \cubbiw\ and its coefficients
\abici\ and identify the independent parameters. We have: the
generalized glueball $\t S$; the degree $n-1$ polynomial $F_0(x)$,
with $n$ coefficients; the polynomial $\t F_2(x^2)$, which has
$(n+1)/2$ coefficients; $\t F_1(x^2)$ which has $(n-1)/2$
coefficients. However, by making use of the first anomaly equation
in Appendix A, one can show that the coefficients of $\t F_1(x^2)$
can be recast as combinations of coefficients of $F_0(x)$, so they
are not free parameters. We are left with a total of
$1+n+(n+1)/2={3\over2}(n+1)$ parameters, which is precisely the
number of vacua, i.e. the low energy SQCDs in the generic vacuum
\genbre. However, it might seem this is not in agreement with the
number of independent deformations of the curve, which has genus
$g=3(n+1)-2$. But recall that the coefficients \abici\ of the
curve are even functions, namely the curve has the automorphism
$x\to-x$ that halves the number of moduli: this means that the
periods of $R(x)$ around $A_i$ and $\hat A_i$ are respectively the
same as those around $A_i'$ and $\hat A_i'$. Finally, the
coefficient of the leading term of the quantum deformation
$F_0(x)$ is fixed as the sum of all the glueballs, just as in the
one--adjoint case we discussed above.

\newsec{The mesons}

We will solve now for the generator of the $X$--dressed mesons.
Its classical expression depends on which of the three classical
phases we are considering: it vanishes in the pseudoconfining
phase, while it is given by \meshig\ in the abelian and nonabelian
higgs phase. Our strategy is again to consider a variation of the
fundamentals and get an anomalous Ward identity.

In our case \treesupal\ the meson deformation is just
$X$--dependent, $m(x)=m_1+m_2 x$. Let us focus then on the
$X$--dependent variation $\delta Q^f=Q^f{1\over x-X}$, which gives
the usual anomaly equation $[m(x)M(x)]_-=\hbar R(x)$, where we
suppressed flavor indices. These considerations still hold if we
consider the generalized meson generators $M_k(x)=\t Q{1\over
x-X}Y^kQ$, whose anomaly equations are $[m(x)M_k(x)]_-=\hbar
R_k(x)$. The explicit solution depends on the vacuum we are
considering. We have three cases
\item{1.} {\it Pseudoconfining branch}. In this case the classical
meson generator vanishes on the first sheet $M(x)|_{cl}=0$, so we
require that the spurious poles coming from the zeroes of $m(x)$
be cancelled
\eqn\pseumes{ M(x)=\hbar\left({R(x)\over m(x)}-{R^I(x_h)\over
m'(x_h)}{1\over x-x_h}\right). }
\item{2.} {\it Abelian higgs branch}. This is characterized by the
meson generator having a pole at the higgs eigenvalue $x_h$ on the
first sheet, whose residue is computed according to the first
expression in \meshig. If we recall the expression of the
resolvent on the second sheet in \semialli, we see that the
boundary conditions are that the meson generator be regular on the
second sheet
\eqn\mesabig{ M(x)=\hbar\left({R(x)\over m(x)}-{R^{II}(x_h)\over
m'(x_h)}{1\over x-x_h}\right). }
This means that we can connect the branches \pseumes\ and
\mesabig\ by moving the pole at $x_h$ from the first to the second
sheet by passing through one of the abelian cuts $I_i$.
\item{3.} {\it Nonabelian higgs branch}. This phase is
characterized by the meson generator having a pole at $x_h$ on the
first sheet, whose residue is computed according to the second
expression in \meshig. If we recall the expression of the
resolvent on the third sheet in \semialli, we see that the
boundary conditions in this case are that the meson generator be
regular on the third sheet
\eqn\mesnabig{ M(x)=\hbar\left({R(x)\over m(x)}-{R^{III}(x_h)\over
m'(x_h)}{1\over x-x_h}\right). }
This means that we can connect the branches \mesabig\ and
\mesnabig\ by moving the pole at $x_h$ from the second to the
third sheet by passing through one of the cuts $I'_i$ around
$x=-a_i$, where $a_i$ is one of the abelian pseudoconfining vacua.
We can connect the nonabelian higgs solution to the
pseudoconfining one by passing the pole from the third to the
first sheet through one of the nonabelian cuts $\widehat
I_i$.\foot{If the meson deformation only depends on the $X$
adjoint, as in \treesupal, then we can solve for the meson
generator $M(x)$, but we can't write a closed equation for the
generator of the $Y$--dressed mesons $M(y)$. In fact, by $\delta
Q^f=Q^f{1\over x-X}{1\over y-Y}$ we get the anomaly equation
$[m(x)M(x,y)]_-=\hbar Z(x,y)$, where $M(x,y)$ and $Z(x,y)$ are
defined in \chiralgen. In the pseudoconfining branch, the mesons
vanish classically, so we expect that the residues of $M(x,y)$
around the poles of $m(x)$ be vanishing in the classical limit.
This gives $M(y)=-\hbar\sum_kZ(x_k,y)/m'(x_k)$. The same reasoning
applies in the case the meson deformation only depends on $Y$
instead, i.e. $\t Q m(Y) Q$. Here, we can solve for the generator
$M(y)$ but we can't get a nice expression for the generator
$M(x)$. Eventually, if the meson deformation depends on both $X$
and $Y$, then there is no easy way to study either meson
generators. }

\noindent We can summarize the solution for $M_k(x)$ in \gemete\
as
 \eqn\diffes{
 M_k(x)={\hbar\over m(x)}\left(R_k(x)-R_k^{(i)}(x_h)\right), }
where $x_h=-m_1/m_2$ and the index $i=I,II,III$ labels
respectively the pseudoconfining, abelian and nonabelian higgs
branches and gives the resolvent on the three different sheets by
\solut. The explicit expressions for the higher $R_k(x)$ are given
in Appendix A. We can characterize the three branches as follows
\eqn\tabmes{\matrix{
       {\rm branch}     \qquad  & \qquad{M_k(x)} \cr&\cr
{pseudoconfining}    \qquad&\qquad  {\rm regular\, on\, I\,
sheet}\cr {abelian\,\, higgs}    \qquad&\qquad  {\rm regular\,
on\, II\, sheet}\cr {nonabelian\,\, higgs}    \qquad&\qquad  {\rm
regular\, on\, III\, sheet} }}

\subsec{Konishi Anomaly}

Let us check that the mesons satisfy the Konishi anomaly in each
low energy SQCD \konishi. To get the expectation values of the
meson operators in one low energy SQCDs, we just integrate the
meson generator $M(x)$ around the corresponding cut. We have to
distinguish whether the SQCD we flow to is in the abelian or in
the nonabelian pseudoconfining vacuum. The $i$--th SQCD coming
from the abelian vacuum \Xveval\ has $N_f$ flavors and
$$
\bigvev{\t QX^jQ}|_i=\oint_{A_i}x^jM(x),
$$
where we suppressed the flavor indices. The first meson gives the
usual Konishi anomaly equation $\t Q_f Q^f=\hbar N_f S_i/m(a_i)$,
where $m(a_i)$ is the effective quarks mass in the $i$--th SQCD.

The $j$--th SQCD that comes from the nonabelian vacuum \Xvevtwo\
gets twice as many flavors and requires some additional
considerations. We can parameterize the $2N_f$ fundamentals as $
(\t Q_\a)_{\t f}^{\pm}$ and $(Q^\a)^{\pm,f}$ where
$f=1,\ldots,N_f$ is the flavor index and $\alpha=1,\ldots,\hat
N_j$ is the color index and the additional index $\pm$ is another
flavor index that comes from the splitting of the color indices in
the nonabelian vacuum and the fact that the rank of the gauge
group is halved in this vacuum. The superpotential of this SQCD is
$W_{eff}=m(\hat a_j)\t Q^+_{f}Q^{+,f}+m(-\hat a_j)\t Q^-_{f}
Q^{-,f}$, so the two type of fundamentals $Q^{+}$ and $Q^-$ have
different mass. We have two different kind of mesons, which are
decoupled. The $+$ mesons are given by the $\hat A_j$ periods, the
$-$ mesons by the $\hat A_j'$ periods
\eqn\mesna{ \bigvev{\t Q^+X^lQ^{+}}|_j= \oint_{\hat
A_j}x^lM(x),\qquad \bigvev{\t Q^-X^lQ^{-}}|_j= \oint_{\hat
A_j'}x^lM(x). }
If we recall the definition of the glueball in this vacuum
\nabeglu, we find that the Konishi anomaly takes two expressions
$\t Q^+ Q^{+}=\hbar N_f\hat S_j/m(\hat a_j)$ and $\t Q^-
Q^{-}=\hbar N_f\hat S_j/m(-\hat a_j)$.

\newsec{Interpolating Between the Three Branches}

In this section we will solve for $T(x)=\Tr{1\over x-X}$ and study
its analytic behavior. We will find that it has only simple poles
with integer residues and that the pseudoconfining, abelian higgs
and nonabelian higgs branches of the gauge theory are described by
three different configurations of the simple poles on the curve.
As a result, the gauge theory curve is a degree three algebraic
curve with marked points. Let us recall what happens in the one
adjoint case \CSW. In that case, the location of the poles of
$T(x)$ characterizes either of the two branches of the theory: a
pole on the physical sheet signals a semiclassical higgs branch,
while when $T(x)$ is regular on the physical sheet we are in a
semiclassical pseudoconfining branch. One can interpolate
continuously between the two branches by moving the pole of $T(x)$
between the first and the second sheet through the branch cuts. In
our two adjoint case, when $T(x)$ is regular on the physical sheet
the theory is in the semiclassical pseudoconfining phase, while
more complicated configurations describe the two higgs branches.
Once again, we can reach all three branches by moving poles around
the three sheets of the Riemann surface \cubbiw.

\subsec{Solving for $T(x)$}

To compute the exact expression of $T(x)$ is rather tedious. The
strategy is analogous to the one we used for the resolvent $R(x)$,
namely we collect some anomaly equations for $U(x,y)$, defined in
\chiralgen, Then we consider its Laurent expansion and extract a
linear equation for $T(x)$. The interested reader will find the
details in Appendix A. By the tree level superpotential
\treesupal\ and the meson deformation $m(x)=m_1+m_2\,x$ we
obtain\foot{The only case one can solve the anomaly equations
explicitly is when $m(x)$ is just linear in $x$ as in \treesupal.
If higher Yukawa couplings are present this procedure does not
work any more.}
\eqn\reshig{ T(x)={N(x)+\delta N(x)\over D(x)}. }
The notation is the following
\eqn\notii{\eqalign{
N(x)=&x^2C_0(x)\left[V'(x)-V'(-x)\right]-\hbar
R(x)\left[C_0(x)+C_0(-x)\right]-x^2\left[2\l\t C_2(x^2)+\a\t
C_1(x^2)\right],\cr D(x)=&\left[x^2 V(x)+{\a^2\over4\l}-2x^2\hbar
R(x)\right] \left[V'(x)-V'(-x)-2\hbar
R(x)\right]-\hbar^2x^2R(x)^2\cr &+\hbar
x^2\left[F_0(x)+F_0(-x)\right]+\hbar\a\t S, }}
and we have introduced the degree $n-1$ polynomials
\eqn\Ck{C_k(x)=\bigvev{\Tr{V'(x)-V'(X)\over x-X}Y^k},}
which are analogous to \delte, but do not vanish classically. In
particular, the leading term of the first polynomial $C_0(x)$ is
the rank of the unbroken gauge group $N_c=C_{0(n-1)}/t_n$. We have
also used the combinations $2x^2\t C_1(x^2)=x[C_1(x)-C_1(-x)]$ and
$2\t C_2(x^2)=C_2(x)+C_2(-x)$. The term $\delta N(x)$ depends on
the meson generators
\eqn\numenew{\eqalign{ \delta
N(x)=&m_2\Bigl(-M(x)[V'(x)-V'(-x)]+\lambda[M_2(x)+M_2(-x)]+{\alpha\over2x}[
M_1(x)-M_1(-x)]\cr&+\hbar R(x)[M(x)+M(-x)]\Bigr), }}
where $M_k(x)=\tQ{1\over x-X}Y^kQ$ are the meson generators whose
exact expression in the quantum theory is given in \diffes. Note
that the term \numenew\ vanishes if the superpotential does not
have Yukawa couplings between the fundamentals and the adjoints.
In the massive theory where $m(X)=const$ we would have just
$T(x)=N(x)/D(x)$. The explicit expression of $\delta N(x)$ depends
on the semiclassical branches through the meson generators.

Once we have the explicit solution \reshig\ we can study its
semiclassical expansion, its asymptotics and its singularities.
The semiclassical expansion on the physical sheet is
\eqn\resT{T(x)={x^2C_0(x)\over x^2 V'(x)+{\alpha^2\over4\lambda}}
+\l x{2\t C_2(x^2)+\a\t C_1(x^2)\over 2v_-(x^2)\left[x^2
V'(x)+{\alpha^2\over4\lambda}\right]}+\hbar{N_cS\over t_n
x^{n+2}}+{\cal O}(\hbar^2). }
The first quantum correction has the asymptotics $x^{-n-2}$, so we
keep it since it contributes to $\langle\Tr X^{n+1}\rangle$. The
higher quantum corrections ${\cal O}(\hbar^2)$ begin at
$x^{2n+3}$, so they do not contribute to the expectation values of
the nontrivial operators in the chiral ring \chicl\ and we can
drop them. With some work one can compute also the asymptotics on
the second and third sheet and find that the resolvent $T(x)$ has
a simple pole at infinity on the first sheet with residue $-N_c$
and a simple pole at infinity on the third sheet with residue
$N_c-2$, where the $-2$ comes from $\delta N(x)/D(x)$. It is
regular at infinity on the second sheet.

Let us find out the other poles of $T(x)$ on the various sheets in
the three different branches. The only singularities of $T(x)$ are
the ones at infinity and the simple poles at the images of the
point $x_h$. The branches enter in the expression \reshig\ of
$T(x)$ only through the meson generators in \numenew. There is a
nice pictorial way to see the three branches. Let us denote by a
cross ``$\times$'' a simple pole with residue $-1$ and with a dot
``$\bullet$'' a simple pole with residue $1$. We can summarize the
singularities in the three branches as\foot{The appearance of a
pole with residue $-1$ for $T(x)$ might seem unexpected. Consider
the semiclassical expansion of $T^{II}(x)$ in \reshig\ on the 2nd
sheet: $D(x)$, $N(x)$ and $\delta N(x)$ are all even functions of
$x$ as $T^{II}(x)$, which is regular at infinity. The contour
integral of $T^{II}(x)$ on a large contour henceforth vanish. We
can close the contour around the finite singularities, which are
the poles at $x_h$ and $-x_h$ and the periods around $A_i$ and
$A_i'$. We have
$\sum_{i=1}^{n+2}\left(\oint_{A_i}+\oint_{A_i'}\right)T^{II}(x)=0$
and  $\left(\oint_{x_h}+\oint_{-x_h}\right)T^{II}(x)=0$. The
residue around $-x_h$ is thus the opposite of the residue around
$x_h$.}
 \eqn\tabtipic{\quad\matrix{
 \matrix{\cr\cr{\rm III\,\,  sheet}\cr{\rm II\,\,  sheet}\cr{\rm I\,\,
 sheet}}\hskip0.5cm
 &\matrix{{\rm pseudo}&\hskip-0.35cm {\rm conf.}\cr
 -x_h & x_h\cr
 \bullet     & \bullet   \cr \times
 & \bullet \cr 0      & 0 }&\hskip1cm
 &\matrix{{\rm abel.}&\hskip-0.1cm {\rm higgs}\cr-x_h & x_h\cr
 0     & \bullet   \cr 0
 & 0 \cr 0      & \bullet }&\hskip1cm
 &\matrix{{\rm nonab.}&\hskip-0.3cm {\rm higgs}&\cr-x_h & x_h\cr
 0     & 0   \cr \times
 & \bullet \cr \bullet      & \bullet }}}
For each branch, in the first column we collect the residues at
the images of $-x_h$ on the three sheets and in the second column
the residues at the images of $x_h$. Note that, just as the meson
generator in \tabmes, each branch is characterized by the
generator being regular on one of the three sheets. Therefore, we
can label the second sheet the {\it abelian higgs sheet} and the
third one the {\it nonabelian higgs sheet}. When the resolvents
are regular on the physical sheet we are of course in the
pseudoconfining phase, as shown in Fig. 2.

\ifig\pseudofig{The pseudoconfining phase. The black and white
dots represent poles for $T(x)$ with residue respectively one and
minus one. The contours $\widehat C'_{nab}$ and $D_{ab}$ enclose
the images of the higgs eigenvalue $-x_h$, while $\hat C_{nab}$
and $C_{ab}$ enclose the images of $x_h$.}
{\epsfxsize=0,8\hsize\epsfbox{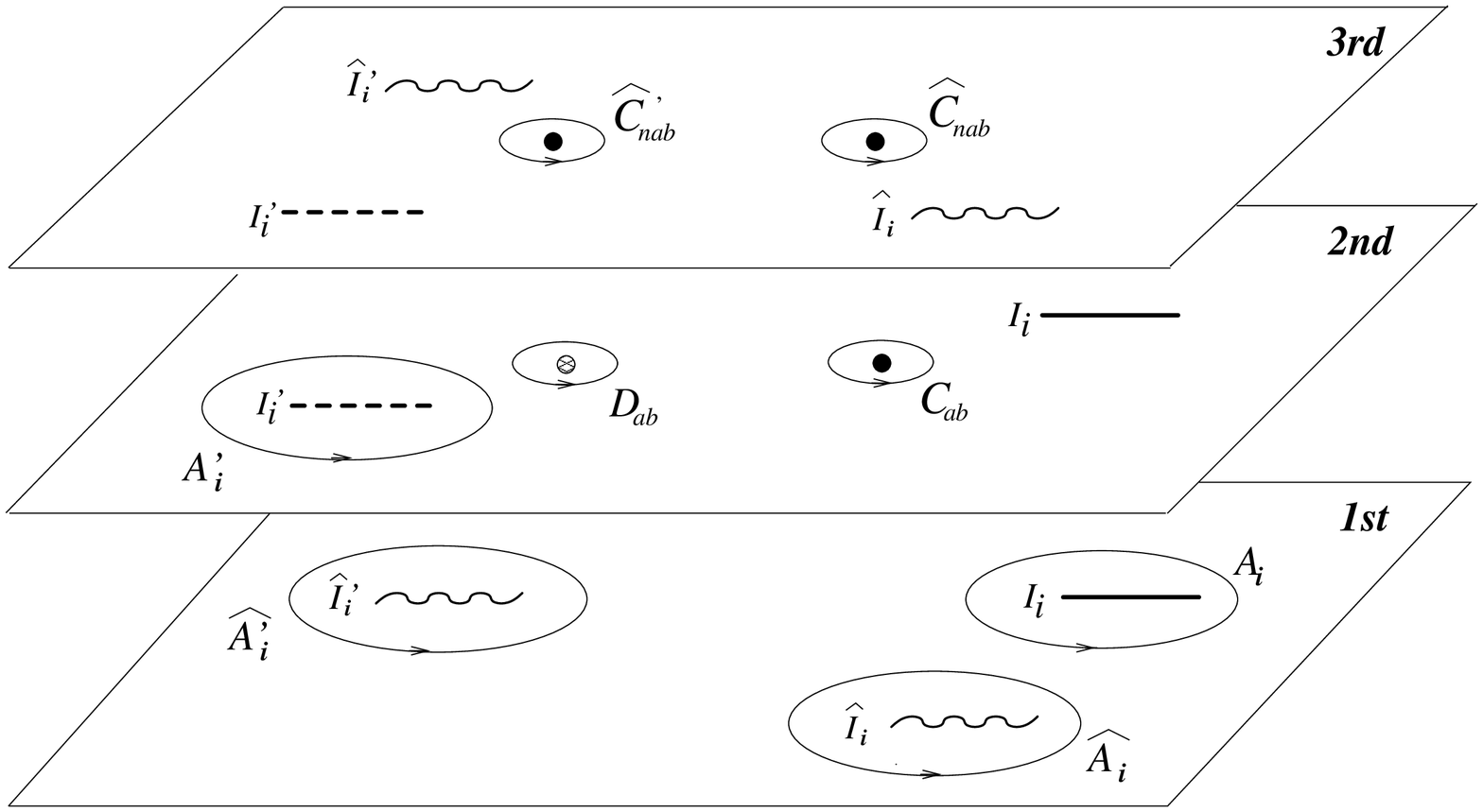}}

Let us consider the $A$--periods of $T$, recalling the definition
of the glueballs in Section 3.3. In the one--adjoint theory \CDSW,
the $A$--periods of $T$ define the ranks of the $i$--th low energy
SQCD as $N_i=\oint_{A_i}T(x)$, but in this case we have two
different kinds of SQCDs, by the flows in the abelian and
nonabelian vacua. The $A_i$ periods define the ranks of the SQCD
in the abelian vacua \Xveval\ as usual
$$ N_i=\oint_{A_i}T(x), $$
while the ranks of the SQCD in the nonabelian blocks \Xvevtwo\ is
computed by either periods around the nonabelian cuts
\eqn\nabiti{ \hat N_i=\oint_{\hat A_i}T(x)=\oint_{\hat A_i'}T(x).
}
With these definitions we recover the residue of $T(x)$ at
infinity in the physical sheet as the sum of the the ranks plus
the higgs poles
$N_c=\sum_{i=1}^{n+2}N_i+2\sum_{i=1}^{n-1\over2}+r_{ab}+2r_{nab}$,
where $r_{ab}$ is 1 in the abelian higgs branch and vanishes
otherwise, while $r_{nab}$ is 1 in the nonabelian higgs branch and
vanishes otherwise.

\subsec{Interpolating Between the Three Phases}

Looking at the table \tabtipic, we can check that the sum of all
residues of $T(x)$ on the curve vanishes. Moreover, when a cross
meets a dot, they annihilate and, viceversa, from a vanishing
residue we can create a pair cross--dot: $\times+\bullet=0$. Now
we can picture the way we interpolate between the three different
branches as follows.
\item{i)}{\it pseudoconfining $\leftrightarrow$ abelian higgs}:
Start with the pseudoconfining phase and move the dot $\bullet$
from the 2nd sheet to the 1st through the cut $I_i$. Due to the
automorphism $x\to-x$, the other cross $\times$ in the second
sheet moves through the symmetric cut $I'_i$ from the 2nd to the
3rd sheet. Once on the 3rd sheet, the cross $\times$ annihilates
with the $\bullet$, being both residues of a pole at $-x_h$, and
we are left with the abelian higgs phase.
 $$
 \quad\matrix{
 \matrix{\cr\cr{\rm III\,\,  sheet}\cr{\rm II\,\,  sheet}\cr{\rm I\,\,
 sheet}}\hskip0.5cm
 &\matrix{{\rm pseudo}&\hskip-0.35cm {\rm conf.}\cr
 -x_h & x_h\cr
 \bullet     & \bullet   \cr \times\uparrow I_i'
 & \bullet\downarrow I_i \cr 0      & 0 }&\qquad
 \matrix{\longrightarrow\cr{}\cr{}\cr{}\cr{}\cr}\qquad
 &\matrix{{\rm abel.}&\hskip-0.1cm {\rm higgs}\cr-x_h & x_h\cr
 \times+\bullet=0     & \bullet   \cr 0
 & 0 \cr 0      & \bullet }}
 $$
When passing the pole through the $i$--th abelian cut $I_i$,  the
rank $N_i$ of the corresponding $i$--th SQCD decreases by one.
This is depicted in Fig.3. The new contour in fact is
$A_i|_{new}=A_i-C_{ab}$ and we find
\eqn\decre{ N_i'=\oint_{A_i}T(x)-\oint_{C_{ab}}T(x)=N_i-1. }
\ifig\pseuhiggs{Fig. 3: Interpolating between the pseudoconfining
and the abelian higgs phase. Start in the pseudoconfining phase in
\pseudofig. Then move the pole $D_{ab}$ to the 3rd sheet through
the cut $I_i'$ and the pole $C_{ab}$ to the 1st sheet through the
cut $I_i$. On the 3rd sheet, the contours $D_{ab}$ and $\hat
C_{nab}'$ combine giving vanishing residue for $T(x)$ at $-x_h$ on
the 3rd sheet. The new period of $T(x)$ on the first sheet is
around the contour $A_i|_{new}=A_i-C_{ab}$ and we find \decre.}
{\epsfxsize=0,8\hsize\epsfbox{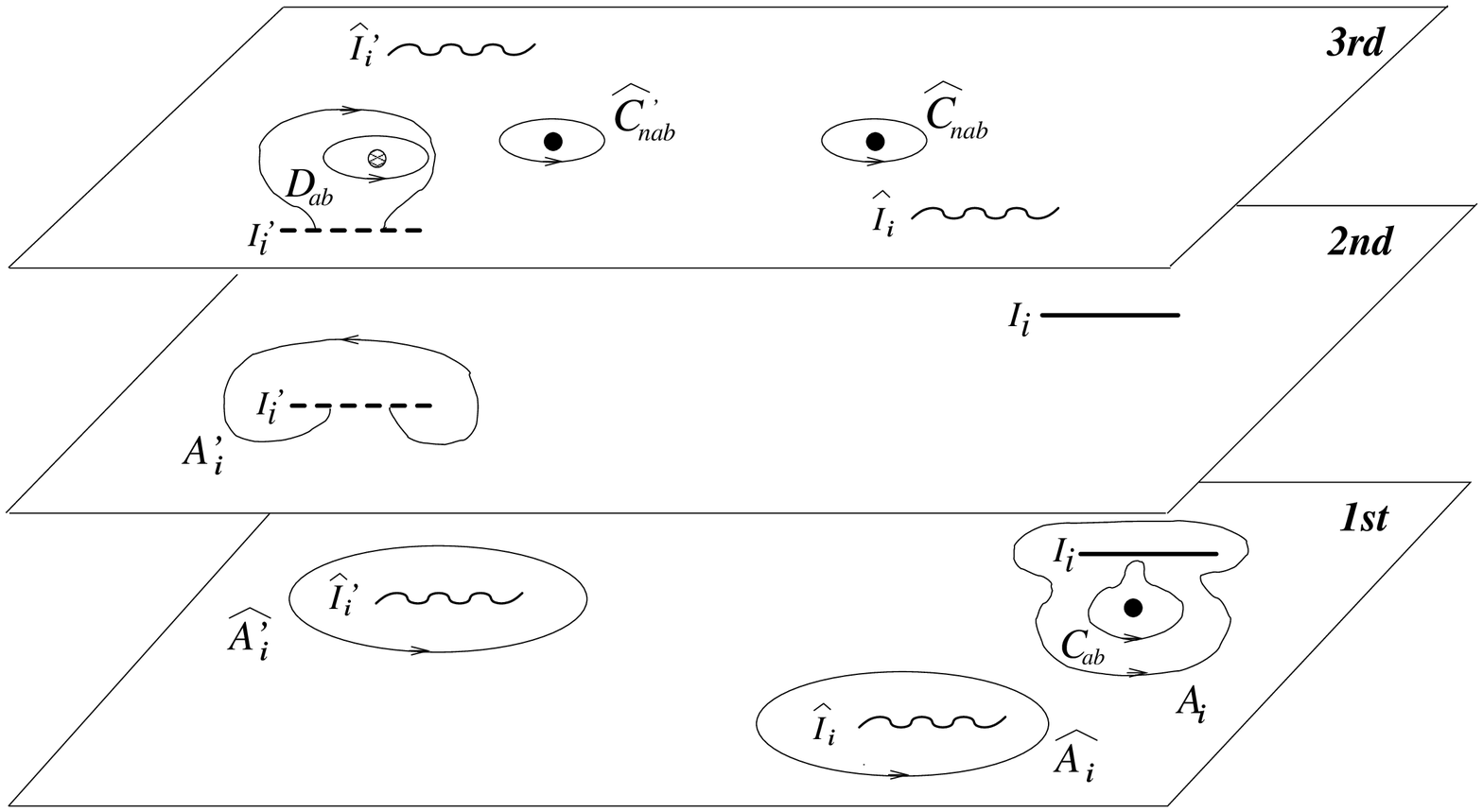}}
\item{ii)}{\it pseudoconfining $\leftrightarrow$ nonabelian
higgs}: Start with the pseudoconfining phase and move the two dots
$\bullet$ from the 3rd sheet to the 1st through the nonabelian
cuts: the pole at $-x_h$ moves through $\widehat I'_i$ and the
pole at $x_h$ moves through $\widehat I_i$.
 $$
 \quad\matrix{
 \matrix{\cr\cr{\rm III\,\,  sheet}\cr{\rm II\,\,  sheet}\cr{\rm I\,\,
 sheet}}\hskip0.5cm
 &\matrix{{\rm pseudo}&\hskip-0.35cm {\rm conf.}\cr
 -x_h & x_h\cr
 \bullet \uparrow\hat I_i'    & \bullet\uparrow \hat I_i   \cr \times
 & \bullet \cr 0      & 0 }&\qquad
 \matrix{\longrightarrow\cr{}\cr{}\cr{}\cr{}\cr}\qquad
 &\matrix{{\rm nonab.}&\hskip-0.1cm {\rm higgs}\cr-x_h & x_h\cr
 0     & 0\cr \times
 & \bullet \cr \bullet      & \bullet }}
 $$
When passing the pole through the $i$--th nonabelian cut $\hat
I_i$, the rank $\hat N_i$ of the corresponding $i$--th SQCD
decreases by one. The new cycles are in fact $\hat A_i|_{new}=\hat
A_i-\hat C_{nab}$ and $\hat A_i'|_{new}=\hat A_i'-\hat C_{nab}'$
and we find
$$
\hat N_i'=\oint_{\hat A_i}T(x)-\oint_{\hat C_{nab}}T(x)=\hat
N_i-1,
$$
or equivalently for the other period around $\hat A_i'$.

\item{iii)}{\it nonabelian higgs $\leftrightarrow$ abelian higgs}:
Start with the nonabelian higgs phase, pass to the pseudoconfining
phase by moving the poles from the physical sheet to the third one
and then move to the abelian higgs phase.

\subsec{Truncation of the Chiral Ring}

The classical chiral ring of the gauge theory \treesupal\ is very
different depending on whether $n=\deg\, V'(x)$ is odd or even. In
the case $n=$ odd, we showed for $V'(x)=x^n$ that the independent
operators are the ones in \chicl, and this still holds when deform
the theory by lower relevant operators as for generic $V'(x)$. In
particular, the chiral ring is characterized by the
following relations, which are the anomaly equations that we
derived in Appendix A and we used to solve for $T(x)$
\eqn\ra{ \Tr X^{k+1}Y+{\alpha\over2\lambda}\Tr X^{k}=0, } \eqn\rb{
\Tr X^l\left(V'(X)+\lambda Y^2\right)Y^k={\cal O}(\hbar), }
\eqn\rc{ \Tr X^{2l+1}Y^{k+2}+{\alpha\over2\lambda}\Tr
X^{2l}Y^{k+1}={\cal O}(\hbar), }
for $k,l\geq 0$. The first classical relation is exact, while the
last two get quantum corrections. One can use these relations to
prove that the operators $\Tr X^{2l+1}Y^2$, higher powers of $\Tr
X^{2l\geq n}Y^2$, $\Tr Y^{j\geq3}$ and $\Tr X^{j\geq n+1}$ can be
expressed in terms of the basis \chicl.

If $n'$ is even, on the other hand, the chiral ring is not
truncated. This would mean that there are an infinite number of
independent operators in the classical chiral ring. We want to
address the $n'$ even case. We start at the IR fixed point $\hat
D$ of the theory with superpotential $\Tr XY^2$, in the notations
of \IW. Then we have two possibilities. If we consider the flow
triggered by the relevant deformation $V'(X)=t_{n'}X^{n'}$ with
$n'=2m$, we can easily see that the chiral ring is not truncated.
But we can consider the different flow, in the bottom line
\eqn\flowo{ \matrix{ {\bf \hat D}
&\longrightarrow&\longrightarrow&\longrightarrow&{\bf
n'}=even\quad{\rm not\,\, trunc.}\cr
                &\searrow       &               &               &\cr
                &               &{\bf n}=odd\quad{\rm trunc.}   &\longrightarrow&{\rm new}\,\,{\bf
                n'}=even\quad{\rm trunc.}
}}
by first turning on a deformation such that $V'(X)=t_{n}X^{n}$
with $n=2m+1$ and flow to the fixed point where the chiral ring is
truncated. Then, we can switch on another relevant deformation
such that $\delta V'(X)=t_{n'}X^{n'}$ with $n'=2m<n$, that
triggers a flow to another fixed point corresponding to the even
case, but this fixed point is different from the untruncated one,
namely here the chiral ring, inherited by the odd case, is still
truncated. We can see this by considering the superpotential
$V'(X)=t_{n}X^{n}+t_{n'}X^{n'}$. At the second fixed point, the
first coupling $t_{n}$ can be set to one, while the coupling
$t_{n'}$ becomes marginal, so that actually
$V'(X)=X^n+t_{n'}X^{n'}$.\foot{The crucial point here is that if
we started directly with the even coupling $t_{n'}X^{n'}$, we
could not use \rc\ to eliminate $\Tr X^{n'}Y^2$. If we start with
the odd coupling, on the contrary, we can do the job and then,
when flowing to the even case, this equation is still valid, by
just setting $t_{n}=1$ and keeping the marginal coupling
$t_{n'}$.} Therefore, the chiral ring for $n'$ even that we get by
flowing down from $n >n'$ contains the following
operators\foot{The analogous computation in the offshell chiral
ring gives the following ${3\over2}n+3$ nontrivial operators $\Tr
W_\alpha^2 X^j$ for $j=0,\ldots,n'+1$, $\Tr W_\alpha^2 Y$ and $\Tr
W_\alpha^2X^{2i}Y^2$ for $j=1,\ldots,{n'\over2}$.}
 $$\eqalign{ \Tr X^j,\qquad j=2,\ldots,n'+1,\cr \Tr
 X^{2i}Y^2,\qquad i=1,\ldots,{n'\over2}. }
 $$

We consider now the relations in the chiral ring of the theory
with no fundamentals and we explain how the classical relations
are modified in the quantum theory. Recall that in the one adjoint
theory with superpotential $W=\Tr V(X)$, the classical chiral ring
is described by the relation $\bigvev{\Tr V'(X)}=0$. In the
quantum theory, the anomaly equation $\langle\Tr{V'(X)\over
x-X}\rangle=2\hbar R(x)T(x)$ tells us that classical relation
still holds, but it gets modified by quantum corrections as long
as we insert higher powers of $X$ as $\bigvev{\Tr X^k V'(X)}={\cal
O}(\hbar)$ if $k\ge1$.

Consider now the theory with two adjoints and superpotential
\treesupal. The classical chiral ring is described by the set of
relations \ra--\rc. The first and the third relations \ra\ and
\rc\ correspond to the $Y$ equation of motion with $X$ insertions
only and with both $X$ and $Y$ insertions. The second relation
\rb\ corresponds to the $X$ equation of motion with both $X$ and
$Y$ insertions. With the help of the anomaly equations for the
$T(x)$ resolvent, that we worked out in Appendix A.3, we can
compute the modifications that these relations are subject to in
the quantum theory. The first relation \ra\ is exact. The second
relation still holds for $l=0$ and gets quantum corrections for
$\l\ge1$ and the third relation \rc\ gets quantum corrections for
$k,l\geq 0$. Therefore we draw the general lesson that, when
considering vacuum expectation values of gauge invariant chiral
operators, the equations of motion a chiral superfield $\Phi$
still hold in the quantum theory, as long as we do not have
additional insertions of $\Phi$ itself in the single trace
correlator.

\newsec{The Classically Invisible Sheets and the Branches}

As anticipated in the Introduction, we would like to get some
general understanding of the curve $\Sigma$ of the ${\cal N}=1$
supersymmetric gauge theory from this analysis of the different
branches of SQCD. Consider a supersymmetric gauge theory with a
matter content such that, once we fix the number of unbroken
$U(1)$s in the low energy theory, there is no order parameter to
distinguish between the various classical vacua in an invariant
way, so have preferred the word {\it branches} rather than phases.
This means that, even if the classical theory has different kinds
of solution to the equations of motion, in the quantum theory we
can reach all the different semiclassical behaviors, with the same
number of unbroken $U(1)$ gauge groups, by continuously moving the
couplings along their moduli space, as in the case of a theory
with matter in the fundamental representation.\foot{Vacua with a
different number of unbroken $U(1)$s, however, describe two
different phases of the theory. In fact, as suggested in \CSW\ and
discussed in \masaki\ for the one adjoint theory, if the $i$--th
low energy SQCD has $N_i=1$, it is not possible to further pass
any pole through the corresponding cut in the onshell theory.} It
is clear that the different branches can only make sense in the
limit of
 \item{i)} large expectation values, which is the semiclassical
 approximation;
 \item{ii)} well separated branch cuts, that is far from singular
 points in the moduli space.

The observables that characterize the different branches are the
ones in the chiral ring of the onshell theory. For instance, in
the particular example of matter in the fundamental and adjoint
representation, they are the resolvents $M(x)$ and $T(x)$, that we
defined in \gemete. The semiclassical branches are then
characterized by the analytic properties of these resolvents on
the curve $\Sigma$, that is by their poles and the respective
residues. We have proposed that an ${\cal N}=1$ gauge theory with
a mass gap is described by a degree $k$ algebraic curve, where $k$
is the number of different branches of the theory. The curve is a
$k$--sheeted covering of the plane, where each sheet corresponds
to a different branch.\foot{We are excluding the case in which
there is also a Coulomb branch, as it happens in the one adjoint
theory \treead\ when $n=N$. In this case, for instance, the gauge
group is broken to its Cartan subalgebra, there is no mass gap
and, in the limit of vanishing superpotential, we recover the
${\cal N}=2$ SQCD.}

In this way we can explain the appearance, in the quantum theory,
of the ``classically invisible sheets'', to quote \CSW. Let us see
how this works and focus the attention on the meson operator
$M(x)$. In our SQCD, the mesons are dressed by the adjoints, but
with a more general matter content they would be dressed in some
other ways and the general picture would not change. Each branch
is characterized by a set of classical expectation values for the
matter fields, which are set to the solutions to the equations of
motion. In our case \treesupal\ for instance we have the
pseudoconfining, abelian higgs and nonabelian higgs branches. Each
branch is defined by the poles and residues of $M(x)$ on the first
sheet at large expectation values (semiclassical regime). By the
generalized Konishi anomaly equations, it follows that the generic
form of $M(x)$ in the branch $A$ is given by
 \eqn\genim{
 M^A(x)=\hbar g(x)R(x)+q^A(x),
 }
where $g(x)$ is a rational function of the couplings, $R(x)$ is
the gauge theory resolvent (which defines the curve) and $q^A(x)$
is another rational function that sets the boundary conditions on
the meson operator, its poles and residues. Only in the classical
limit do the branches make sense, thus we are interested in just
the last term $q^A(x)$. In general, this depends on the resolvent
$R(p_i)$ evaluated at the poles $p_i$, which are the images of the
classical higgs expectation values. Since we assumed that there is
no invariant way to distinguish the branches, we can connect all
of them by changing continuously the boundary conditions $q^A(x)$,
that is by moving the poles $p_i$ between the sheets through the
branch cuts. Now, since the resolvent $R(p_i)$ on the curve
$\Sigma$ gets as many different classical limits as the number of
sheets (this is the way we identify the sheets), it turns out
that, when taking the classical limit of large poles in the first
sheet, we obtain as many different expressions as the number of
sheets. Each one of them is a solution to the equations of motion
and therefore a different branch. Suppose that we have $k$
branches but $k+1$ sheets. Then, we can continue the pole $p_i$ to
that extra sheet and compute the classical limit for $M(x)$ on the
first sheet, but this corresponds to a new solution to the
equations of motion and so we have found a new branch.

For a generic ${\cal N}=1$ gauge theory, this holds with the
following two caveats:
 \item{--} The number of sheets corresponds to the number of branches that we can
distinguish in the effective description we are using. In our case
of the deformed $D_{n+2}$ theory \treesupal, in the $X$ effective
description we can see only three branches, but we will argue
below that new branches could be identified in the $Y$ effective
description.
 \item{--} If there is an order parameter that characterizes one
 phase in an invariant way, then it seems plausible that the
 corresponding sheet be disconnected.

\subsec{An Example: the Branches of SQCD}

Let us see how our proposal works in the paradigmatic case of
SQCD, where we have now a complete picture of all the possible
branches. Depending on the extra matter content we can test our
conjecture in  different situations.

\vskip0.3cm \noindent{\it Ordinary SQCD: One Sheet}

Consider SQCD with gauge group $U(N_c)$. We can describe the
offshell curve of this theory in a confining vacuum by adding a
massive adjoint superfield $X$ and integrating it out. The tree
level superpotential is
 $$
 W_{SQCD}={t_1\over2}\Tr X^2+m\t Q Q,
 $$
with $t_1\gg m$. This theory classically has only one branch, the
pseudoconfining one, in which both $X$ and the fundamentals
vanish. The corresponding curve is
 \eqn\sqcdcu{
 y^2=t_1^2x^2+4\hbar t_1 S,
 }
where $S$ is the glueball. This looks like a double cover of the
$x$ plane with two branch points at $a^\pm=\pm2\sqrt{S/t_1}$. But
it is just a fake covering and the curve \sqcdcu\ actually
describes the Riemann sphere. We have just one sheet corresponding
to the one classical branch.

\vskip0.3cm \noindent{\it SQCD with One Adjoint: Two Sheets}

Consider $U(N_c)$ SQCD with one adjoint $X$ and a confining phase
superpotential
 $$
 W=\Tr V(X)+\t Q m(X)Q,
 $$
where $V'(x)$ has degree $n$ and $m(x)$ degree $n-1$. This theory
has received a lot of attention. For $n<N$, it has two branches,
the pseudoconfining and the (abelian) higgs one. The curve is the
well known hyperelliptic Riemann surface $y^2=V'(x)^2+\hbar f(x)$,
where the degree $n-1$ polynomial $f(x)$ is the quantum
deformation. For $n>1$, this is a genuine double--sheeted covering
of the $x$ plane. As explained in \CSW, we can continuously
interpolate between the pseudoconfining and higgs branch by moving
the poles of $M(x)$ and $T(x)$ from the second to the first sheet.
The first sheet corresponds to the pseudoconfining branch and the
second to the higgs branch.

\vskip0.3cm \noindent{\it SQCD with Two Adjoints: Three Sheets}

This is the theory \treesupal\ that we have discussed at length.
Classically, it has three branches: pseudoconfining
\Xveval--\Xvevtwo, abelian higgs \quahi\ and nonabelian higgs
branch \nonhi. The curve \cubbiw\ is a three--sheeted covering of
the plane and each sheet corresponds to a different branch, as
explained in \tabtipic. \vskip0.3cm

Up to now we have considered the effective description of two
adjoint SQCD when integrating out the adjoint superfield $X$.
However, we can as well integrate out the other adjoint $Y$ and
study the effective theory encoded in the resolvent
$$
R(y)=\thipi\bigvev{\Tr{\Wsquare\over y-Y}}.
$$
To find an algebraic equations for $R(y)$ we can use the same
anomaly equations that we used for $R(x)$, but Laurent expand them
in inverse powers of $x$ instead, the details are in Appendix A.
One finds that $R(y)$ satisfies a degree $2^n$ algebraic equation
\ferrari. The curve in the $Y$ effective description is thus a
$2^n$ sheeted covering of the plane. This might seem weird at
first, since we conjectured that each sheet of the gauge theory
curve is related to a different semiclassical phase. However, the
three branches of our gauge theory are just the ones that we
obtain by coupling the fundamentals with the adjoint $X$, and are
the ones we can study in the $X$ effective description of the
theory. It would be nice to check that, when adding the most
generic meson deformation $\delta W=\t Qm(X,Y)Q$ to the
superpotential, more complicated vacua appear corresponding to new
higgs solutions, that we can just tell from each other in the
effective $Y$ description. On the $X$ side they would be
undistinguishable from the three branches we already considered.

\newsec{The Magnetic Dual}

In this Section we will present some results on the Seiberg dual
of the theory \treesupala. A Seiberg dual description of the
theory with gauge group $SU(N_c)$ and superpotential
\eqn\elscf{ W_{el}=t_n \Tr X^{n+1}+\l\Tr XY^2+\beta \Tr X, }
has been proposed by Brodie in \brodie.\foot{This theory has no
Lagrange multiplier for $Y$. Setting $\a=0$ means that the adjoint
$Y$ gains an overall nonvanishing $U(1)$ component $\Tr Y$.} This
theory flows to a nontrivial fixed point in the infrared. The
magnetic dual is an ${\cal N}=1$ $SU(3nN_f-N_c)$ gauge theory with
two adjoint chiral superfields $\t X$ and $\t Y$, $N_f$ magnetic
fundamentals $\t q^{\t f}$ and $q_f$ and $3n$ gauge singlets
$(P_{l,j})^f_{\t f}$ for $l=1,\ldots,n$ and $j=1,2,3$, each of
which transforms in the $(N_f,\bar N_f)$ of the flavor symmetry
group. A magnetic superpotential was proposed
\eqn\scfsu{ W_{m}=\bar t_n \Tr \t X^{n+1}+\bar \l\Tr \t X\t Y^2+
\t q \t m(X,Y) q+\t\beta\Tr \t X, }
where $\bar t_n=-t_n$ and $\bar\l=-\l$ and $\t m(X,Y)$ is a
polynomial that couples the magnetic fundamentals to the gauge
singlets and the adjoints. This represents a Legendre transform
between the electric and magnetic mesons. As opposed to our theory
with confining phase superpotential \treesupal, the theory \elscf\
has just $n+2$ one--dimensional vacua and no two--dimensional
vacua is present in this case.

We are interested in finding its generalization when we deform it
by the confining phase superpotential
\eqn\genel{ W_{el}=\Tr V(X)+\lambda\Tr XY^2+\beta\Tr X+\alpha\Tr
Y, }
allowing for abelian as well as and nonabelian vacua (the latter
are not present in Brodie's case).\foot{In this section we will
explicitly keep track of the multiplier $\beta$ by displaying it
separately from the adjoint polynomial $V'(x)=\sum_{i=1}^n t_i
x^i$ on the electric as well as on the magnetic side.} The
magnetic tree level superpotential is this case is not just the
analogue of \treesupala, but contains an extra term
\eqn\genma{ W_{mag}=\Tr \bar V(\t X)+\bar \lambda\Tr \t X\t
Y^2+\bar s\Tr \t Y^2+\t \beta\Tr \t X+\bar\alpha\Tr \t Y+\t q\t
m(X,Y)q, }
corresponding to the extra coupling $\bar s$, which is fixed by
duality to $\bar s=\l{t_{n-1}\over t_n}{N_f\over \bar N_c}$.
Unfortunately, for generic $n$ one cannot solve the anomaly
equations in the magnetic theory on a closed set of resolvents,
due to this extra $\bar s$ coupling, so we have to stick to the
classical duality map. However, in the $D_3$ case we can integrate
out one of the two adjoints and flow to the well known one adjoint
duality, which was studied by Kutasov, Schwimmer and Seiberg \KSS.
In this case, we will be able to give the duality map in the
quantum theory between the electric and magnetic couplings and
quantum deformation.  As a consistency check, we reobtain the
duality map which was found in \mio.

Let us make a brief digression about the magnetic polynomial $\t
m(x,y)$. The {\it rationale} behind this Legendre transform term
is that it must decouple the electric and magnetic mesons in
different low energy SQCD blocks. This superpotential term can be
conveniently parameterized by the kernel
\eqn\magpot{ \t m(\t X,\t Y)={1\over\mu^4}\oint dz\,dw\,{\bar
V'(z)-\bar V'(\t X)\over z-\t X} {w(\bar \lambda w^2+\t \beta)-\t
Y(\bar \lambda \t Y^2+\t \beta)\over w-\t Y}P(z,w), }
that reproduces the one--adjoint term of \KSS\mio\ in the case
$n=1$. Note that, for $n>1$, \magpot\ works only in the case
$v_+(x^2)=0$ and for $\alpha=0$. In \magpot\ we collected the $3n$
gauge singlets $(P_j)^{f}_{\t f}$ into a single meromorphic
function
\eqn\psin{ P(z,w)={P^{(1)}(z)\over w}+{P^{(2)}(z)\over
w^2}+{P^{(3)}(z)\over w^3},\qquad
P^{(j)}(z)=\sum_{l=1}^n{P_l^{(j)}\over z^l}. }
For nonvanishing $\alpha$, \magpot\ does not decouple all the low
energy SQCD blocks. The difficulty in finding such polynomial in
this generic case might be related to the fact that, as we
stressed in Section 2.3, we have to impose also the D--term
equations of motion \dcond\ on the pseudoconfining vacua.

Consider the vacua of the magnetic theory \genma, they are very
similar to the electric ones \Xveval\ and \Xvevtwo. The equations
of motion are
\eqn\mageom{ \bar V'(\t X)+\t\b+\bar\l Y^2=0,\qquad \{\bar\l \t
X+\bar s,\t Y\}+\t\a=0, }
and their irreps are still one--dimensional vacua and
two--dimensional vacua. The $n+2$ abelian vacua are analogous to
\Xveval\ but with magnetic eigenvalues and multiplicities instead,
such that \onepola\ is replaced by $\bar p(x)=\left(x^2+{\bar
s\over\bar\l}\right)[\bar V'(x)+\t\b]+{\t\a^2\over4\bar\l}$ and
$\bar b_i=-{\t\a\over2 \bar\l( \bar a_i+{\bar s\over\bar\l})}$.
The $(n-1)/2$ nonabelian vacua are analogous to \Xvevtwo, but each
block of the adjoint is replaced by $\t X=\hat{\bar
a}_i\sigma_3-{\bar s\over\bar \l}\Id_2$ and $\t Y=\bar
d_i\sigma_3+\bar c_i\sigma_1$ and $\hat{\bar a}_i,\bar d_i,\bar
c_i$ are fixed by \mageom.

We can use the SQCD duality relation $\bar N_i=\#\,{\rm
flavors}-N_i$ in each low energy SQCD block to check that the
ranks of the gauge groups match. The electric low energy theory we
flow to in the $n+2$ abelian vacua is SQCD with $N_f$ flavors,
while in the ${n-1\over2}$ nonabelian vacua it is SQCD with $2N_f$
flavors. Therefore we have
$$\bar N_c=\sum_{i=1}^{n+2}\bar
N_i+2\sum_{i=1}^{n-1\over2}\hat{\bar N}_i
=\sum_{i=1}^{n+2}(N_f-N_i)+2\sum_{i=1}^{n-1\over2}(2N_f-\hat N_i)=
3nN_f-N_c.
$$

Now we want to find the classical duality map. We will use a
strategy analogous to KSS \KSS. Let us first try a naive map
between electric and magnetic eigenvalues $\bar a_i=a_i$ and $\bar
b_i=b_i$.\foot{Since we will stick to the electric pseudoconfining
phase, we can forget about the magnetic polynomial $\t m(X,Y)$ for
the moment.} We have to impose the tracelessness condition on both
the electric and magnetic adjoint vacua. On the electric side, the
nonabelian vacua \Xvevtwo\ are already traceless and only the
abelian vacua contribute, fixing the Lagrange multipliers $\beta$
and $\alpha$. On the magnetic side, the condition $\Tr \t Y=0$
gives $\sum_{i=1}^{n+2}{1\over a_i}=0$, but this is automatically
satisfied because the linear term in the abelian polynomial
\onepola\ vanishes. Then we have to impose $\Tr \t X=0$, where
also a contribution from the nonabelian magnetic vacua appear.
Since $a_i$ are the roots of \onepola\ we have
$\sum_{i=1}^{n+2}a_i=-{t_{n-1}\over t_n}$ and the tracelessness
condition can not be satisfied, unless $t_{n-1}=\bar s=0$. If
$t_{n-1}$ vanishes, then the trivial map works.

\subsec{The Shift of the Electric and Magnetic Theory}

To find the map for nonvanishing $t_{n-1}$ we follow the usual
trick in singularity theory and shift the electric and magnetic
adjoints. The new feature is that we need to add the new coupling
$\Tr \t Y^2$ to the magnetic side. Then we impose the
tracelessness conditions on the shifted adjoints and find that the
naive map works in the shifted variables.

Consider the electric theory \genel\ and shift $X$ as
$X=X_s-B\Id$n. We do not shift $Y$ since duality would fix the $Y$
shift to zero. Then the electric superpotential reads
\eqn\genles{\eqalign{ W_{el}=&\Tr V_s (X_s)+\beta_s\left(\Tr
X_s-BN_c\right)+\lambda\Tr X_sY^2+ \alpha\Tr Y-\lambda B\Tr
Y^2+\phi N_c, }}
where
\eqn\curci{\eqalign{ V_s(X_s)=&\sum_{i=1}^n{g_i\over
i+1}X_s^{i+1},\qquad g_l=\sum_{i=l}^n\pmatrix{i\cr
l}t_i(-B)^{i-l},\cr
\beta_s=&\beta+\sum_{i=1}^nt_i(-B)^i,\qquad\phi=\sum_{i=1}^n{i+2\over
i+1}t_i(-B)^{i+1}. }}
The shifted equations of motion are $V_s'(X_s)+\beta_s+\lambda
Y^2=0$ and $\lambda\{X_s-B,Y\}+\alpha=0$. Their irreps are still
abelian and nonabelian pseudoconfining vacua. Then, by imposing
the tracelessness conditions $\Tr \t X_s=\Tr Y=0$ we fix the
electric Lagrange multipliers.

On the magnetic side we have to consider the magnetic
superpotential \genma\ with the extra deformation $\delta\bar
W=\bar s\Tr \t Y^2$. At the end this new coupling will be fixed by
duality as a function of the other couplings. Let us shift the
magnetic theory as $\t X=\t X_s-\bar B\Id$, the superpotential
reads
$$\eqalign{ \bar W_{m}=&\Tr \bar V_s (\t X_s)+\t \beta_s\left(\Tr
\t X_s-\bar B\bar N_c\right)+\bar \lambda\Tr \t X_s\t Y^2+
\t\alpha\Tr \t Y+(\bar s-\bar\lambda \bar B)\Tr \t Y^2+\bar \phi
\bar N_c+f_s, }
$$
where the notation is as in \curci\ but with magnetic quantities
instead and $f_s(t_l,\l,B)$ is the shifted coupling dependent
function. If we solve the equations of motion we still find the
abelian and nonabelian vacua and, by imposing again that the
magnetic adjoints be traceless we find
\eqn\tracema{ \Tr\t X_s=\sum_{i=1}^{n+2}(N_f-N_i)(\bar a_{s,i})
+2\left(\bar B-{\bar
s\over\bar\l}\right)\sum_{i=1}^{n-1\over2}(2N_f-\hat N_i)=0, }
where $\bar a_{s,i}$ are the shifted magnetic abelian vacua. Once
we impose both electric and magnetic tracelessness conditions, we
can postulate the naive match of the shifted eigenvalues
\eqn\naimat{ \bar a_{s,i}=a_{s,i}, }
and we crucially drop the dependence upon the vacua inside
\tracema\ by fixing the shifts
\eqn\naimatti{\eqalign{ B&={g_{n-1}\over 2(n-1)g_n},\cr \bar
B&=B+{\bar s\over\bar\l}. }}
By comparing the electric and magnetic abelian vacua we get the
map between the shifted couplings and Lagrange multipliers
\eqn\mapshi{ \eqalign{ \bar g_l=-g_l,\qquad \bar\l=-\l,\cr \bar
\beta_s=-\beta_s,\qquad \bar \alpha=-\alpha. }}

A sufficient condition for the map between the operators to be
independent of the vacua is that the electric and magnetic
superpotential match and the coupling dependent function
$f_s(t_l,\l,B)$ do not depend on the vacuum. One can easily check
this last requirement by differentiating the effective action with
respect to the shifted couplings $ \Tr X_s^{l+1}=-\Tr \tilde
X_s^{l+1}+(l+1){\partial f_s\over\partial g_l}$ and see that
$\partial_{g_l}f_s$ does not depend on $N_i,\hat N_i$.

\subsec{The Map in the Original Couplings}

By using \naimat, \naimatti\ and the map \mapshi\ we can reobtain
the relation between the eigenvalues and the couplings in the
original parametrization of the theory
\eqn\mapppi{\eqalign{ \bar a_i&=a_i-{\bar s\over \bar\l},\cr \bar
t_l&=-\sum_{i=l}^n\pmatrix{i\cr l}t_l\left({\bar s\over
\bar\l}\right)^{i-l}. }}
While the abelian $X$ eigenvalues are shifted by the duality, the
nonabelian eigenvalues match exactly, as well as the $Y$
expectation values
\eqn\mapptwo{ \eqalign{ X:&\qquad\widehat{\o{a}}_i=\hat a_i,\cr
Y:&\qquad\bar b_i=b_i,\quad \bar c_i=c_i,\quad \bar d_i=d_i. }}
By imposing the tracelessness condition in the unshifted magnetic
adjoint we fix $\bar s$ in terms of the other couplings of the
theory
$$\bar s=\l{t_{n-1}\over t_n}{N_f\over\o N_c}. $$
In other words, we can write down the electric superpotential as
\genel, while the magnetic one \genma\ reads in electric variables
\eqn\magella{ W_{m}=-\Tr V(\tilde X+{\bar
s\over\bar\l})-\l\Tr(\tilde X+{\bar s\over\bar\l})\t
Y^2-\alpha\Tr\t Y-\beta\Tr\t X+f(coupl.). }

\subsec{Duality for $D_3$: the Quantum Theory}

We would like to solve for the chiral ring operator \chiralgen\
both in the electric and magnetic side in the quantum theory and
find the map between the dual quantum deformations $F_k(x)$. For
generic superpotentials \genel\ and \genma\ this seems impossible,
since on the magnetic side the anomaly equations do not close any
more on an algebraic equation for the magnetic resolvent $\t
R(x)$, due to the extra coupling $\Tr \t Y^2$. The only case in
which we can solve both electric and magnetic quantum theories is
when the adjoint polynomial is just a mass term
\eqn\dtre{ W_{el}=\Tr\left({t_1\over2} X^2+\beta  X+\lambda
XY^2+\alpha Y\right)+m\t QQ. }
We will denote \dtre\ the $D_3$ theory. The duality map in this
case will reproduce exactly the KSS duality in the case of a
one--adjoint SQCD \KSS\mio, that is we will see that $D_3\sim
A_3$, as expected from singularity theory. The $D_3$ theory have
the $n+2=3$ abelian vacua \Xveval\ but no nonabelian vacua
\Xvevtwo, which are only present if $n\geq3$. Here $X$ is massive
and we can integrate it out upon its equations of motion,
obtaining, at energies below the mass scale $t_1$, an effective
superpotential $U_{eff}(Y)=-{1\over 2t_1}\Tr(\beta +\lambda
Y^2)^2+\alpha\Tr Y$, whose derivative is
\eqn\uprime{ U'(y)=-{2\lambda\over t_1}(\beta y+\lambda
y^3)+\alpha. }
In the quantum theory, the anomaly equation for the resolvent
$R^Y(y)$ reduces to the hyperelliptic curve
\eqn\resymas{ \hbar^2R(y)^2-U'(y)R(y)-{1\over4}f(y)=0. }
This is the usual anomaly equation for the one matrix model \CDSW,
where $U'(y)$ is the effective superpotential \uprime\ and
\eqn\noneu{f(y)={8\l\over t_1^2}(\beta F_0+\l F_2+\l yt_1\t S+\l
y^2F_0),}
is the quantum deformation, that we expressed in terms of the
parameters that we used in the solution for $R(x)$ in
\abici.\foot{From the algebraic equation in Appendix A we get
$f(y)={8 \lambda\over t_1} (\lambda y^2S+\lambda y\t S-t_1\hat
S_1)=0$, where $\hat S_1=\thipi\bigvev{\Tr\Wsquare X}$. By using
the recursion relations in Appendix A we can also express the
parameter $\hat S_1$ in terms of the parameters $F_0=t_n S, F_2$
that we used in the solution for $R(x)$ in \abici\ as $t_1\hat
S_1=-{2\lambda\over t_1}F_2-\beta  S$.} The solution of the
anomaly equation \resymas\ is the well known $2\hbar
R(y)=U'(y)-\sqrt{U'(y)^2+\hbar f(y)}$. The physical picture in
this case is the following. Classically, the resolvent $R(y)$ has
three poles located at the classical vacua $y=b_i$, where the
$b_i$'s are given in \Xveval. These are the roots of the cubic
effective polynomial $U'(y)$. In the quantum theory, each pole
splits into two branch points, that connect the first and the
second sheet of the hyperelliptic curve. In this case we just have
two classical branches, the pseudoconfining and the abelian higgs
vacua. The nonabelian higgs phase is only present if $n\geq3$. So
we just have two sheets as in the one--adjoint theory. We can get
the expectation values of the dressed mesons in the quantum theory
by computing the contour integrals of the meson generator $M(y)=\t
Q{1\over y-Y}Q=m^{-1}R(y)$ and find
\eqn\mesisi{ \t QQ={S\over m},\qquad \t QYQ={\t S\over m},\qquad
\t QY^2Q={F_2\over t_1}. }

The magnetic theory dual to \dtre\ is
$$
W_{mag}={\bar t_1\over2}\Tr \t X^2+\bar\l\Tr \t X\t
Y^2+\t\beta\Tr\t X+\t\a\Tr\t Y+\t q\t m(X,Y) q+\bar m\tr
P_{1}^{(1)},
$$
where the last term corresponds to the electric mass term for the
fundamentals and the trace is over flavor indices. The magnetic
polynomial $\tilde m(X,Y)$ in this case is \magpot, even if we
keep a nonvanishing $\a$ multiplier. Actually, the polynomial $\t
m(X,Y)$ is equivalent to what we get by using the effective
quartic one--adjoint polynomial $\bar U(\t Y)$ in \uprime. In
fact, the $\t X$ dependence drops and we are left with
\eqn\magmas{\t m(X,Y)={\bar t_1\over\mu^4}\oint dw\,{\bar
U'(w)-\bar U'(\t Y)\over w-\t Y},}
The magnetic singlet equations of motion are
\eqn\singlem{\t q Y^jq=-\delta_{j,2}{\bar m\mu^4\over\bar \l \bar
t_1}.}

The anomaly equations for the resolvents $\t R^{\t X}(x)$ and $\t
R^{\t Y}(y)$ are the same as in the electric theory, \cubbiw\ and
\resymas. We want to solve for the singlets in the magnetic
theory, following the method in \mio. The magnetic meson generator
$\t M(y)=\t q{1\over y-\t Y}q$ satisfies the anomaly equation $[\t
m(y)\t M(y)]_-=\t R^{\t Y}(y)$. The generic solution is
\eqn\solmesni{ \t M(y)={\t R^{\t Y}(y)\over \t m(y)}+{\t r(y)\over
\t m(y)}. }
The way we solve it is by first fixing $\t r(y)$ to cancel
spurious singularities from the zeros of $\t m(y)$ and then
solving for $\t m(y)$ such that the singlet equations of motion
\singlem\ are satisfied. The solution in the pseudoconfining phase
is $\t r(y)=0$ and
\eqn\tildras{ \t m(y)=-{\bar t_1^2\over \bar \l}{\t f(y)\over
8\bar m\mu^4}, }
where $\t f(y)={8\bar \l\over \bar t_1^2}(\t\beta \t F_0+\bar\l \t
F_2+\bar\l y \bar t_1\t{\bar S}+\bar \l y^2\t F_0)$ is the quantum
deformation of $\t R^{\t Y}(y)$ and $\t F_k$ are the quantum
deformations of $\t R^{\t X}(x)$, the magnetic version of \abici.
Formally it is the same expression as in \noneu, but replacing
electric with magnetic quantities.

The magnetic polynomial in terms of the singlets reads $\t
m(y)={\bar t_1\over\mu^4}(\bar\l P_1^{(3)}+\bar \l y
P^{(2)}_1+(\t\beta+\bar\l y^2) P_1^{(1)})$. We can read off the
expression of the quantum expectation value of the gauge singlets
in terms of the quantum deformations
\eqn\threegau{ P_1^{(1)}=-{\t F_0\over \bar m\bar t_1},\qquad
P_1^{(2)}=-{\t{\bar S}\over \bar m}, \qquad P^{(3)}=-{\t F_2\over
\bar m\bar t_1}. }
If we match them directly to the electric mesons \mesisi\ we find
the map
duality map in the quantum theory between the quantum deformations
$\t F_0=F_0$ and $\t F_2=F_2$ and then between the couplings, the
Lagrange multipliers and the glueballs
\eqn\mappuno{\eqalign{ \bar t_1=&-t_1,\qquad \bar \l=-\l,\qquad
\t\beta=-\beta,\cr \bar S=&-S,\qquad \bar{ \t S}=-\t S. }}

\newsec{Further Directions}

In this paper we have proposed a general explanation of the
presence of the classically ``invisible'' sheets in the curves of
${\cal N}=1$ supersymmetric gauge theories. In general, the gauge
theory curve is realized as a $k$--sheeted covering of the plane.
One of these sheets is visible in the classical theory, while the
remaining sheets are not accessible semiclassically bu only in the
full quantum theory. A convenient method to compute this curve is
by the DV prescription, that relies on the planar limit of a
related matrix model or, correspondingly, on solving a set of
anomaly equations in the gauge theory. We considered theories with
matter content such that, once we fix the number of unbroken
$U(1)$ gauge groups, there is no order parameter to distinguish
the various classical vacua, hence we denoted the different kinds
of classical solutions as branches. Our proposal is that, under
these circumstances, there is a one to one correspondence between
the number of branches and the degree of the curve.

This proposal holds trivially in the case of ordinary SQCD and has
been verified also for SQCD with one adjoint chiral superfield in
\CSW. In this paper, we have worked out the classical and quantum
theory of SQCD with two adjoints and superpotential \treesupal\
and we have verified that the proposal works also in this case. In
particular, we have shown that this theory has three classical
vacua, namely the pseudoconfining, the abelian higgs and the
nonabelian higgs ones. We have proven that in the quantum theory
we can associate each sheet of the cubic curve to each of these
three branches by looking at the singularities of some meromorphic
functions on the curve. Moreover, we have argued that one can
interpolate continuously between all the classical vacua with the
same number of unbroken $U(1)$ factors. It would be interesting to
verify our conjecture for other gauge theories with a higher
degree DV curve, in particular one can address the following
cases.

Consider a $U(N_c)$ gauge theory with one adjoint and an
additional chiral superfield in the symmetric (or antisymmetric)
representation. Its DV curve is a cubic, as in our two adjoint
SQCD, and has been computed in \klemm\naccu. One could couple this
theory to matter in the fundamental representation and find out
the classical branches. According to our proposal, we expect to
see, in addition to the pseudoconfining vacua of \klemm\naccu, two
different higgs vacua and, in the quantum theory, we expect the
three branches (with the same number of unbroken $U(1)$s) to be
connected continuously.

The second theory is a quiver $SU(N_c)\times SU(N_c)$ gauge theory
with matter in the bifundamental representation. The curve of this
theory is again a cubic \naccu, but it has a weird feature, namely
each node of the quiver sees a particular sheet as its own
physical sheet and the leftover sheet seems mysterious. It would
be interesting to add fundamental matter to this theory and
classify its classical branches, then study the quantum theory and
see how we can connect the different branches by moving the poles
between the sheets. In this way one could clarify our proposal in
the case of a quiver theory. Moreover, a Seiberg dual theory to
this quiver with fundamentals has been discussed in \products. It
would be nice to see the dual description of the electric branches
on the DV curve, which is the same for both dual pairs, along the
lines of \mio.

\vskip0.2cm {\it The $E_n$ type SQCD}\vskip0.2cm

Finally, an extremely interesting theory where to test our
proposal is SQCD with two adjoints and $E_n$ type (according to
the ADE classification of \IW) tree level superpotential. For
instance, one can consider the $E_6$ theory with superpotential
$W=\Tr Y^3+ \Tr X^4$ deformed by lower dimensional operators. The
classical vacua of this theories are not known. However, by
studying their flows in connection with the $a$ theorem, \IW\
argued that there are an infinite number of irreps of the
equations of motion with vanishing fundamentals (which we called
the pseudoconfining branch). First of all, it would be nice to see
explicitly whether the number of pseudoconfining vacua is actually
infinite. One could find also the higgs vacua and classify all the
branches of the theory, then compute the ${\cal N}=1$ curve and
verify if the degree of the curve agrees with the number of
branches.

As a byproduct of this analysis, one would shed light on the
following mystery. The analytic structure of an ${\cal N}=1$ curve
is such that, on the physical sheet, the number of branch cuts are
in correspondence with the classical pseudoconfining vacua and, in
the classical limit, each branch cut shrinks to a point
corresponding to a pseudoconfining vacuum. In this case, if the
number of pseudoconfining vacua is infinite, it is not at all
clear what the curve would look like, since we would expect an
infinite number of branch cuts on the physical sheet. Moreover,
one could consider the geometric engineering of this theory as a
type $IIB$ superstring theory on a certain local Calabi Yau
threefold, in the framework of \CKV\ferrari.\foot{Some problems
about the engineering of the Yukawa coupling to the fundamentals
have been outlined in \okuda. Anyway one can consider the theory
without fundamentals and discuss the pseudoconfining vacua to
begin with.} The classical theory is described by the geometry of
a $P^1$ bundle over a particular $ALE$ space, which is the
resolution of an $E_n$ singularity. The classical pseudoconfining
vacua of the gauge theory should be seen in the geometry as the
compact holomorphic curves of the threefold. According to the
geometric transition conjecture, in the quantum theory these
holomorphic curves are replaced by three spheres, whose volume is
proportional to the gauge theory glueballs. But if we have an
infinite number of pseudoconfining vacua, as argued by \IW, it is
not at all clear how to make sense of the classical geometric
picture in the first place, whether there are an infinite number
of holomorphic curves in the resolved geometry and, finally, how
to perform the blow down map, if any, and compute the deformed
Calabi Yau.

\vskip0.2cm \centerline{\bf Acknowledgements}

I would like to thank Adam Schwimmer for help and encouragement.
It is also a pleasure to thank M. Bertolini, G. Bonelli, K.
Intriligator, M. Matone, A. Ricco and A. Wulzer for discussions
and especially T. Grava for tricks about cubic curves.

\appendix{A}{Anomaly Equations}

In this Appendix we derive the various anomaly equations we used
above to solve for the resolvents. First we get the cubic equation
\cubbiw\ satisfied by the gauge theory resolvent $R(x)$, following
Ferrari who solved for the planar limit of the related two matrix
model \ferrari. We will show the semiclassical expressions of the
resolvent on the different sheets. Then, we will solve for the
resolvent $T(x)$ in \reshig. Finally, we write down the recursion
relations for the $R(y)$.

\subsec{The curve}

There are many variations that one can try, but only very few of
them are useful. In particular, Ferrari \ferrari\ has shown that
the following three variations
\eqn\variations{\eqalign{ 1st:\quad&\delta X=0,\qquad \delta
Y=\thipi{W_\alpha W^\alpha\over x-X},\cr 2nd:\quad&\delta
X=\thipi{\Wsquare\over x-X}{1\over y-Y} ,\qquad \delta Y=0,\cr
3rd:\quad&\delta X=0,\qquad \delta
Y=\thipi\Wsquare\overx\overy{1\over -x-X}. }}
give the following anomalous Ward identity
\eqn\loopeq{\eqalign{ 1st:\quad&\l R_1(x)=\l{\t S\over
x}-{\alpha\over 2x}R(x),\cr 2nd:\quad&\left[V'(x)-\hbar R(x)+\l
y^2\right]Z(x,y)= \l yR(x)+\l
R_1(x)\cr&\qquad\qquad\qquad\qquad\qquad\qquad\qquad\qquad
\thipi\bigvev{\Tr\Wsquare{V'(x)-V'(X)\over x-X}\overy},\cr
3rd:\quad&Z(x,y)Z(-x,y)=\l\left[R(x)+R(-x)\right]-\left(\l
y+{\a\over2x}\right)Z(x,y)-\left(\l y-{\a\over2x}\right) Z(-x,y),
}}
where $R_k(x)$ are the generalized resolvent in \genres, $Z(x,y)$
is the chiral operator in \chiralgen, and $\t
S=\thipi\langle\Wsquare Y\rangle$. By expanding the loop equations
\loopeq\ in powers of $y$ we find the recursion relations
\eqn\recone{\l R_{k+2}(x)=\left[\hbar
R(x)-V'(x)\right]R_k(x)+F_k(x),} \eqn\reconeo{ \l
\left[R_{q+2}(x)+R_{q+2}(-x)\right]+{\alpha\over
2x}\left[R_{q+1}(x) -R_{q+1}(-x)\right]+\hbar
\sum_{k+k'=q}R_k(x)R_k(-x)=0,}
for $k\geq0$, recalling the definition \delte\ of the quantum
deformations $F_k(x)$. The strategy is to plug \recone\ into
\reconeo\ and get at $k=0$ an equation for $R(-x)$, then at $k=2$
use it to obtain the closed equation in $R(x)$. By introducing $\t
w=\hbar R(x)-V'(x)$ we get the following cubic equation
\eqn\cubico{\t w^3+\t a(x^2) \t w^2+\t b(x^2) \t w+ \t c(x^2)=0, }
where the coefficients are
\eqn\coeffo{\cases{\eqalign{\t
a(x)=V'(x)+V'(-x)-{\a^2\over4\l},}\cr \eqalign{\t
b(x)=V'(x)V'(-x)- {\a^2\over4\l}[V'(x)+V'(-x)]+\hbar
[F_0(x)+F_0(-x)+{\alpha \t S\over x^2}],}\cr \eqalign{\t
c(x)&=-{\a^2\over4\l}V'(x)V'(-x)+\hbar\Bigl(F_0(-x)V'(x)+
F_0(x)V'(-x)\cr&+\l[F_2(x)+F_2(-x) -\hbar{{\t S}^2\over
x^2}]+{\alpha\over2x}\left[{\t S\over x}[V'(x)+V'(-x)]
+F_1(x)-F_1(-x)\right]\Bigr).} }}
Since in the coefficients there appear negative powers of $x$, we
have to rescale the equation \cubico\ by multiplying by $x^2$.
Setting $w=x^2\t w$ we find our cubic equation \cubbiw, where the
new coefficients are \abici.

\subsec{Semiclassical Resolvents}

The semiclassical expansion of the resolvent $R(x)$ on the three
sheets is
\eqn\semiri{\eqalign{ R^I(x)=&{x^2F_0(x)+\alpha\t S/2\over
p(x)}+\l x{2\t F_2(x^2)+\alpha\t F_1(x^2)\over2v_-(x^2)p(x)}+{\cal
O}(\hbar),\cr \hbar R^{II}(x)=&V'(x)+{\alpha^2\over4\lambda
x^2}-\hbar{x^2F_0(x)+\alpha\t S/2\over
p(x)}-\hbar{x^2F_0(-x)+\alpha\t S/2\over p(-x)}\cr &-\hbar\l
x^2{2\t F_2(x^2)-\alpha\t F_1(x^2)\over p(x)p(-x)}+{\cal
O}(\hbar^2),\cr \hbar
R^{III}(x)=&V'(x)-V'(-x)+\hbar{x^2F_0(-x)+{\alpha\t S\over2}\over
p(-x)}-\hbar\l x{2\t F_2(x^2)+\alpha\t F_1(x^2)
\over2v_-(x^2)p(-x)}+{\cal O}(\hbar^2), }}
where $p(x)$ is the polynomial \onepola, whose roots are the
abelian vacua \Xveval, and $v_-(x^2)$, whose roots are the
nonabelian vacua \Xvevtwo, is the odd part of the adjoint
polynomial $V'(x)$. By looking at \semiri\ we get a quick preview
of the structure of the branch points of the gauge theory curve
\cubbiw. Indeed, each pole in the semiclassical resolvent splits
up into two branch points in the full quantum theory. If the
resolvent has a pole at the same value of $x$ on two different
sheets, in the quantum theory a branch cut will appear, connecting
those same sheets. Therefore, the branch cuts $I_i$ coming from
the splitting of the abelian vacua at the roots of $p(x)$ will
connect the first and the second sheet. The branch cuts $I_i'$,
symmetric of the former with respect to the origin, i.e. coming
from the roots of $p(-x)$, connect the second and the third
sheets. The branch cuts $\widehat I_i$ and $\widehat I_i'$ from
the nonabelian vacua, i.e. the roots of $v_-(x^2)$, connect the
first and the third sheets. This confirms the analysis of Section
3.3.

\subsec{The Resolvent $T(x)$}

Consider the theory \treesupal\ with meson deformation
$m(x)=m_1+x\,m_2$. Consider the variations \variations\ but drop
the field strength factor $-{1\over32\pi^2}\Wsquare$ and find the
following three anomaly equations
\eqn\loopt{\eqalign{ 1st:\quad&T_1(x)=-{\alpha\over2\l x}T(x),\cr
2nd:\quad&\left( V'(x)+\l y^2-\hbar R(x)\right)U(x,y)+m_2M(x,y)=
\hbar T(x)Z(x,y)+\cr&\qquad\qquad\qquad\qquad\qquad\qquad\qquad
+\l yT(x)+\l T_1(x) +\bigvev{\Tr{V'(x)-V'(X)\over x-X}\overy},\cr
3rd:\quad&\hbar\left[Z(x,y)U(-x,y)+Z(-x,y)U(x,y)\right]=-\l
y\left[U(x,y)+U(-x,y)\right]\cr
&\qquad\qquad\qquad\qquad\qquad\qquad\qquad-{\alpha\over2x}\left[U(x,y)-U(-x,y)\right]
+\l \left[T(x)+T(-x)\right], }}
where the chiral operators $Z(x,y)$, $U(x,y)$ and $M(x,y)$ are
given in \chiralgen\ and we set to zero the terms proportional to
$u^\a,w^\a$ in the supersymmetric vacuum. $T_k(x)$ are the
generalized resolvents \gemete. The mesons contribute only through
the second anomaly equation with the term proportional to $m_2$.
Recalling the definition \Ck\ of the degree $n-1$ polynomials
$C_k(x)$, we consider the Laurent expansion in powers of $y$ of
the second and third equations in \loopt\ and find the recursion
relations
\eqn\recT{ \l T_{k+2}(x)=[\hbar R(x)-V'(x)]T_k(x)-m_2M_k(x)+\hbar
R_k(x)T(x)+C_k(x),}
\eqn\recTT{\eqalign{
\l\left[T_{k+2}(x)+T_{k+2}(-x)\right]+{\alpha\over2x}\left[T_{k+1}(x)-T_{k+1}(-x)\right]+\cr
+\hbar\sum_{q+q'=k}\left[R_q(x)T_{q'}(-x)+R_q(-x)T_{q'}(x)\right]=0,
}}
for $k\geq0$, where $M_k(x)$ are the generalized meson generators
in \gemete. Notice that these recursion relations are linear in
$T_k$, whereas the recursion relations for $R_k$ in \recone\ and
\reconeo\ are bilinear. Plugging \recT\ into \recTT\ at  $k=0$ we
solve for $T(-x)$, then at $k=2$ we find a linear equation for
$T(x)$, whose solution is precisely
$$
T(x)={N(x)+\delta N(x)\over D(x)},
$$
where $N,\delta N$ and $D$ are given in \notii\ and \numenew.

\subsec{The resolvent $R(y)$}

We would like to solve for the resolvent
$R(y)=\thipi\Tr{\Wsquare\over y-Y}$. There are two methods
\ferrari, but we will use the most intuitive one. Consider the
anomaly equations \loopeq. To solve for $R(x)$ we used their
Laurent expansion in $y$, but now we can use their Laurent
expansion in powers of $x$ and find the following recursion
relations
\eqn\kone{ \sum_{i=0}^nt_iR_i(y)+\lambda y^2R(y)-\lambda
yS-\lambda \t S=0, }
\eqn\resyrec{ \sum_{i=0}^nt_iR_{k+i+1}(y)-\hbar \sum_{i=0}^{k}\hat
S_i R_{k-i}(y) +\lambda y^2 R_{k+1}(y)-\lambda y\hat
S_{k+1}+{\alpha\over2}\hat S_{k}=0, }
\eqn\quadry{ 2\lambda yR_{2k+1}(y)=2\lambda \hat
S_{2k+1}+\hbar\sum_{i=0}^{2k}(-1)^iR_i(y)R_{2k-i}(y)-\alpha
R_{2k}(y), }
for $k\ge0$, where $R_k(y)=\thipi\Tr{\Wsquare\over y-Y}X^k$ are
the generalized resolvents and $\hat S_k=-{1\over 32\pi^2}\Tr
W_\alpha W^\alpha X^k$. If we combine these three equations we get
a closed degree $2^n$ algebraic equation for $R(y)$.

\appendix{B}{Solution of the Cubic Equation}

Consider the cubic equation
\eqn\cuz{w^3+aw^2+bw+c=0.}
First get rid of the subleading term by the shift $w=z-a/3$,
obtaining $z^3+3\gamma z+2\delta=0$, where we introduced
$3\gamma=b-{a^2\over3}$ and $2\delta=c-{ab\over3}+{2\over27}a^3$.
Now the trick is to replace $z$ with two variables under a useful
constraint. Set $z=u+v$ and get
$u^3+v^3+3(u+v)(uv+\gamma)+2\delta=0$. If we just choose
$uv+\gamma=0$ then we get
$$\cases{u^3+v^3+2\delta=0,\cr uv+\gamma=0.}$$
We solve the quadratic equation $u^6+2\delta u^3-\gamma^3=0$,
obtaining $u^3=-\delta+\sqrt{\delta^2+\gamma^3}$. The solutions
for $u$ picks up the three cubic roots of unity, $1$,
$e^{i{2\pi\over 3}}$ and $e^{-i{2\pi\over3}}$, obtaining
$$
\eqalign{u^{(I)}=(-\delta+\sqrt{\delta^2+\gamma^3})^{1\over3},\cr
u^{(II)}=e^{i{2\pi\over 3}}u^{(I)},\cr u^{(III)}=e^{-i{2\pi\over
3}}u^{(I)},}
$$
and analogous solutions for $v=-{\gamma\over u}$. The solutions to
\cuz\ are therefore $w^{(i)}=u^{(i)}-{\gamma\over
u^{(i)}}-{a\over3}$, that we can list
\eqn\zzz{\eqalign{ &
w^{(I)}=(-\delta+\sqrt{\delta^2+\gamma^3})^{1\over3}
-{\gamma\over(-\delta+\sqrt{\delta^2+\gamma^3})^{-{1\over3}}}-{a\over3},\cr
&
w^{(II)}=e^{i{2\over3}\pi}(-\delta+\sqrt{\delta^2+\gamma^3})^{1\over3}
-e^{-i{2\over3}\pi}{\gamma\over(-\delta+\sqrt{\delta^2+\gamma^3})^{-{1\over3}}}
-{a\over3},\cr &
w^{(III)}=e^{-i{2\over3}\pi}(-\delta+\sqrt{\delta^2+\gamma^3})^{1\over3}
-e^{i{2\over3}\pi}{\gamma\over(-\delta+\sqrt{\delta^2+\gamma^3})^{-{1\over3}}}
-{a\over3}. }}

\appendix{C}{Holomorphic Differentials}

In this appendix we compute a basis for the holomorphic
differentials on the curve \cubbiw. We use the method of divisors
in the notations of \farkra. Let us denote by
$$
[g]={P_1^{\alpha_1}\ldots P_n^{\alpha_n}\over Q_1^{\beta_1}\ldots
Q_m^{\beta_m}},
$$
the divisor of $g$, where $P_i$ is a zero of degree $\alpha_i$ and
$Q_j$ is a pole of degree $\beta_j$. The degree of the divisor is
given by $deg[g]=\sum_i\alpha_i-\sum_j\beta_j$. The Riemann--Roch
theorem states the degree of a meromorphic function $g$ is
$deg[g]=0$, while for an abelian differential $\omega$ the degree
is $deg[\omega]=2g-2$. We need to compute the divisors of $dx$,
$x$, $\partial_w f=3(w(x)^2+\gamma(x^2))$ and $w(x)$.

The differential $dx$ vanishes at the branch points and has a
double pole at $\infty$ on each of the three sheets. If we denote
by $\Delta_B$ the divisor corresponding to the $6(n+1)$ branch
points we find
$$
[dx]={\Delta_B\over
P_{\infty_I}^2P_{\infty_{II}}^2P_{\infty_{III}}^2},
$$
so that $deg[dx]=6n=2g-2$. The function $x$ has the following
divisor
$$
[x]={{\cal O}_I{\cal O}_{II}{\cal O}_{III}\over
P_{\infty_I}P_{\infty_{II}}P_{\infty_{III}}},
$$
where ${\cal O}_i$ represents the origin on each sheet.

Now let us consider $\partial_w f$. It vanishes at the branch
point locus $\Delta_B$, while $\partial_w f(x,w)\sim_{x\sim\infty}
x^{2(n+2)}$ on each sheet. Thus, by Riemann--Roch, we are missing
six zeroes. If we study the asymptotics at small $x$ we find that
$\partial_w f_{x,w(x)}\sim x^3$ on the 1st and 3rd sheets while
$\partial_w f_{x,w(x)}\sim const$ on the 2nd sheet, so that
\eqn\divfw{ [\partial_w f]={\Delta_B {\cal O}_I^3{\cal
O}_{III}^3\over
P_{\infty_I}^{2(n+2)}P_{\infty_{II}}^{2(n+2)}P_{\infty_{III}}^{2(n+2)}}.
}

Consider now $w(x)$. We need to study its zeroes for small $x$, in
order to cancel the poles coming from \divfw. For small $w$ we can
approximate the curve \cubbiw\ by $b(x)w+c(x)=0$ so that $w$
vanishes at the roots of $c(x)$ which are not roots of $b(x)$. So
we expect a double zero at $x=0$ and a bunch of $2n$ nonvanishing
other zeroes, whose corresponding divisor we denote by $C_{2n}$.
The asymptotic expansion of the solutions $w(x)$ is
$w(x)\sim_{x\sim0}x^2$ on the 1st and 3rd sheets and
$w(x)\sim_{x\sim0}const$ on the 2nd sheet so that its divisor is
$$
[w]={{\cal O}_I^2{\cal O}_{III}^2C_{2n}\over
P_{\infty_I}^{n+2}P_{\infty_{III}}^{n+2}}.
$$

To build the holomorphic differentials we have to take care of the
poles coming from \divfw\ at the points ${\cal O}_{I,III}$, so
that
\eqn\triple{ {dx\over \partial_w f},\quad {xdx\over \partial_w f},
}
have triple and double poles, respectively, while
\eqn\singles{ {x^2dx\over \partial_w f},\quad {wdx\over \partial_w
f}, }
have just single poles at ${\cal O}_{I},{\cal O}_{III}$. Therefore
we have to eliminate \triple\ but we can take a linear combination
of \singles\ with vanishing residue. Therefore, a basis for the
holomorphic differentials is given by
\eqn\bashol{\eqalign{ {(c_1 x^2+c_2 w)dx\over w^2+\gamma(x)}&,\cr
{x^jdx\over w^2+\gamma(x)}&,\qquad j=3,\ldots,2n+2,\cr
{x^kwdx\over w^2+\gamma(x)}&,\qquad k=1,\ldots,n. }}
We have in total $3n+1=3(n+1)-2=g$ holomorphic differential as
expected.

\listrefs

\end